\documentclass[12pt]{article}
\usepackage{amsmath,amssymb}
\usepackage{graphicx,psfrag,epsf}
\usepackage{enumerate}
\usepackage{natbib}
\usepackage{color,colortbl}
\usepackage{algorithm, algorithmic, graphicx}
\usepackage{comment}
\usepackage{caption}

\newcommand{\blind}{0}

\newtheorem{theorem}{Theorem}
\newtheorem{lemma}{Lemma}

\newtheorem{remark}{Remark}

\newcommand{\reva}[1]{{\color{red} #1}}

\begin{document}

\def\spacingset#1{\renewcommand{\baselinestretch}%
{#1}\small\normalsize} \spacingset{1}


\if0\blind
{
  \title{\bf Group-Orthogonal Subsampling for Hierarchical Data Based on Linear Mixed Models}
  \author{Jiaqing Zhu$$\hspace{.2cm}\\
    Department of Statistics, KLAS and School of Mathematics and Statistics\\
       Northeast Normal University Changchun, China\\
    Lin Wang$$ \\
    Department of Statistics, Purdue University, West Lafayette, IN, USA\\
    and \\
    Fasheng Sun$^{*}$ \\
       Department of Statistics, KLAS and School of Mathematics and Statistics\\
       Northeast Normal University, Changchun, China\\
       $^{*}${Corresponding author}
    }
    \date{}
  \maketitle
}

\if1\blind
{
  \bigskip
  \bigskip
  \bigskip
  \begin{center}
    {\LARGE\bf Group-Orthogonal Subsampling for Hierarchical Data Based on Linear Mixed Models}
\end{center}
  \medskip
} \fi

\bigskip
\begin{abstract}

Hierarchical data analysis is crucial in various fields for making discoveries. The linear mixed model is often used for training hierarchical data, but its parameter estimation is computationally expensive, especially with big data. Subsampling techniques have been developed to address this challenge.
However, most existing subsampling methods assume homogeneous data and do not consider the possible heterogeneity in hierarchical data. To address this limitation, we develop a new approach called group-orthogonal subsampling (GOSS) for selecting informative subsets of hierarchical data that may exhibit heterogeneity.
GOSS selects subdata with balanced data size among groups and combinatorial orthogonality within each group, resulting in subdata that are $D$- and $A$-optimal for building linear mixed models. Estimators of parameters trained on GOSS subdata are consistent and asymptotically normal.
GOSS is shown to be numerically appealing via simulations and a real data application.
Theoretical proofs, R codes, and supplementary numerical results are accessible online as Supplementary Materials.

\end{abstract}

\noindent%

\spacingset{1.45}
\section{Introduction}
\label{sec:intro}

The unprecedented growth of data in modern research poses significant challenges in terms of storage and analysis.
First, an individual's computing resources may not have the capacity to store the entire dataset due to its large size.
Second, even after the dataset has been loaded into memory, traditional analysis methods may be too slow or even impractical due to the large volume of data \citep{bates2014computational,gao2017efficient}.

Subsampling has been widely used to tackle the issue of storage capacity and accelerate data analysis.
Several subsampling techniques have been developed to address the challenges of big data, generally aiming to optimize the downstream modeling.
For example, for linear regression, \cite{ma2015leveraging} proposed to use the leverage score to construct nonuniform subsampling probabilities.
Using the optimal design theory in experimental design, \cite{wang2019information} proposed an information-based optimal subdata selection (IBOSS) method based on the $D$-optimality criterion.
Inspired by the excellent properties of two-level orthogonal arrays under linear models, \cite{WangL} proposed an orthogonal subsampling (OSS) approach and showed that the OSS method typically outperforms existing methods in minimizing the mean squared errors (MSE) of the estimated parameters and maximizing the efficiencies of the selected subdata.
Some other subsampling works for linear regression include \cite{li2020modern}, \cite{ren2021subdata}, \cite{wang2022balanced}, and \cite{yu2022subdata}, among others.
Subsampling methods are also widely studied when other downstream models are considered, for example, the generalized linear model \citep{ai2018optimal}, quantile regression \citep{WangMa,fan2021optimal, ai2021optimal, shao2022optimal}, multiplicative model \citep{ren2022optimal}, nonparametric regression \citep{meng2020more, sun2021asymptotic, meng2022smoothing, zhang2023independence}, Gaussian process modeling \citep{he2022gaussian} and the model-free scenario \citep{mak2018support,shi2021model}.
In addition, \cite{meng2021lowcon} proposed the ``Lowcon" method to address the presence of model misspecification.
\cite{xie2023optimal} proposed an optimal subsampling method for online streaming data.
\cite{yu2022optimal} considered the optimal subsampling method in a distributed environment.
Readers may also refer to \cite{yu2023review} for a comprehensive review of subsampling methodology.

Knowledge discovery in various fields often relies on the analysis of complex data with a hierarchical structure. For example, students could be sampled from within schools, patients from within doctors, medical records from within individuals, or participants in psychological tests from within communities. For more applications, see, for example, \cite{raudenbush1993crossed,mcculloch2004generalized,bennett2007netflix,jiang2007linear,gao2017efficient,gao2019estimation}.
When the covariates of different groups in a dataset come from distinct distributions, they may demonstrate intra-group homogeneity and inter-group heterogeneity. Consequently, selecting a subset of data that has this hierarchical structure requires additional consideration. Existing subsampling methods often assume that the covariates are homogeneous throughout the entire dataset. Using these methods may overlook critical information contained in hierarchical data. Therefore, it is imperative to develop specialized subsampling techniques that can accurately identify and capture the valuable information in such data.


In this paper, we investigate the optimal subsampling method for hierarchical data by assuming that the data points come from a linear mixed model, which allows both fixed and random effects and is particularly used to analyze the data with a hierarchical structure, see \cite{jiang2007linear, gao2019estimation}.
We develop a group-orthogonal subsampling (GOSS) approach to tackle the memory and computational barriers of linear mixed models. GOSS is particularly designed for data with a hierarchical structure and targets two merits of the selected subdata: data size balance among groups and combinatorial orthogonality within each group. First, GOSS achieves data size balance among groups so that all groups contribute equally to the subdata.
Second, GOSS selects the subdata from each group that approximate an orthogonal array (OA) to extract informative data points. OAs are universally optimal and have been employed in subdata selection for first-order linear regression \citep{WangL}. Our first original contribution lies in extending the theory that establishes the optimality of OAs to the context of the linear mixed model. Consequently,
the selected subdata by GOSS is guaranteed to be $D$- and $A$-optimal for the generalized least squares (GLS) estimator of a linear mixed model.
Numerical results in this paper and Supplementary Materials demonstrate that GOSS outperforms existing methods in minimizing the MSE of parameter estimators and the prediction error over the full data.
Regarding the computing time, for a large full data size $N$ with $R$ groups of $p$-dimensional observations and a fixed subdata size $n$, the computational complexity is $O(Np\log (n/R))$, which is a little faster than $O(Np\log n)$ from OSS and as low as $O(Np)$ from IBOSS. In addition, GOSS is naturally suitable for distributed parallel computing to further accelerate the computation.
Theoretical results are provided to show the consistency and asymptotic normality of the GLS estimator obtained on the selected subdata.

The rest of the paper is organized as follows.
Section \ref{sec2} introduces the notations of the linear mixed model and the fundamental framework for the GOSS method.
Section \ref{sec3} introduces the OA and derives their theoretical optimality for obtaining the GLS estimator of a linear mixed model.
Section \ref{sec3.2} proposes the GOSS method and investigates the asymptotic property of the estimator based on the GOSS subdata.
Section \ref{sec4} and Section \ref{sec5} evaluate the GOSS algorithm via simulation studies and a real-world application.
Section \ref{sec6} concludes the paper. Technical proofs and R codes are provided in Supplementary Materials.

\section{The framework}\label{sec2}

Denote the full data as $\{\mathbf{x}_{ij}, y_{ij}\}_{i = 1,\ldots,R}^{j = 1,\ldots,C_i}$, which include $R$ groups and $C_i$ observations in $i$th group for $i=1,\ldots,R$, so that the full data size is $N = \sum_{i=1}^{R}C_i$. Here $\mathbf{x}_{ij}$ is a $p$-vector of covariates for the $j$th unit in the $i$th group, the first component of $\mathbf{x}_{ij}$ is 1, and $y_{ij}$ is its response. Consider the following linear mixed model,
\begin{align}
y_{ij} = \mathbf{x}_{ij}^{T}\boldsymbol{\beta} + a_{i} + e_{ij},    \mathbf{x}_{ij}\in \mathbb{R}^{p \times 1}, i = 1, 2,\ldots,R,  j = 1, 2,\ldots,C_i, \label{1}
\end{align}
where $\boldsymbol{\beta} \in \mathbb{R}^{p \times 1}$ is a vector of fixed effects, $a_i$ is the independent and identically distributed (i.i.d.) random effect associated with the $i$th group, $a_i \sim (0,  \sigma_A^2)$, and $e_{ij}\sim (0, \sigma_E^2)$ is the error term independent from $a_i$.
In the model in \eqref{1}, two observations in the same group are assumed to have constant correlation $\sigma_{A}^2/(\sigma_{A}^2 + \sigma_{E}^2)$, and observations from different groups are uncorrelated. More details about the linear mixed models can be found in \cite{jiang2007linear}.


Let $\mathbf{X} = (\mathbf{X}_{1}^{T}, \ldots, \mathbf{X}_{R}^{T})^T \in \mathbb{R}^{N \times p}$ with $\mathbf{X}_{i} = (\mathbf{x}_{i1},\ldots,\mathbf{x}_{iC_i})^T = (\mathbf{1}_{C_i}, \mathbf{Z}_i)$ and $\mathbf{Z}_i = (\mathbf{z}_{i1},\ldots,\mathbf{z}_{iC_i})^T$ and $\mathbf{Y} = (\mathbf{Y}_{1}^{T}, \ldots, \mathbf{Y}_{R}^{T})^T \in \mathbb{R}^{N \times 1}$ with $\mathbf{Y}_{i} = (y_{i1}, \ldots, y_{iC_i})^T$, for $i=1,\ldots,R$.
The $\mathbf{Z}_{i}$ may be distinctly distributed for different $i$.

We are commonly interested in the estimator of $\boldsymbol{\beta}$, whose GLS estimator based on the full data is given by
\begin{align*}
\hat{\boldsymbol{\beta}} = (\mathbf{X}^T\mathbf{V}^{-1}\mathbf{X})^{-1}\mathbf{X}^T\mathbf{V}^{-1}\mathbf{Y}
\end{align*}
when $\sigma_A^2$ and $\sigma_E^2$ are known, where $\mathbf{V} = \text{Cov}(\mathbf{Y}) = \sigma_E^2\mathbf{I}_N + \sigma_A^2\mathbf{A}$, and $\mathbf{A} \in \mathbb{R}^{N\times N}$ is a block diagonal matrix with the $i$th block $\mathbf{1}_{C_i}\mathbf{1}_{C_i}^T$.
The estimator $\hat{\boldsymbol{\beta}}$ needs $O(Np^2)$ time complexity to calculate, which is not an easy task when $N$ is big. When $\sigma_A^2$ and $\sigma_E^2$ are unknown, they are estimated from data, making the process even slower.

Now consider taking a subset of size $n$ from the full data, where $n_{i}$ points are from the $i$th group so that $n= \sum_{i=1}^{R}n_{i}$. Denote the selected subdata as $\{\mathbf{x}^*_{ij}, y^*_{ij}\}_{i = 1,\ldots,R}^{j = 1,\ldots,n_{i}}$. 
Let $\mathbf{X}^* = (\mathbf{X}_{1}^{*T}, \ldots, \mathbf{X}_{R}^{*T})^T$ with $\mathbf{X}_{i}^{*}= (\mathbf{x}^*_{i1},\ldots,\mathbf{x}^*_{in_i})^T = (\mathbf{1}_{n_i}, \mathbf{Z}^*_i)$ and $\mathbf{Z}^*_i = (\mathbf{z}^*_{i1},\ldots,\mathbf{z}^*_{in_i})^T$, $\mathbf{Y}^* = (\mathbf{Y}_{1}^{*T}, \ldots, \mathbf{Y}_{R}^{*T})^T$ with $\mathbf{Y}^*_{i} = (y^*_{i1}, \ldots, y^*_{in_i})^T$. The GLS estimator based on the subdata is given by
\begin{align}
\hat{\boldsymbol{\beta}}^* = (\mathbf{X}^{*T}\mathbf{V}^{*-1}\mathbf{X}^*)^{-1}\mathbf{X}^{*T}\mathbf{V}^{*-1}\mathbf{Y}^*, \label{subbetahat}
\end{align}
where
$\mathbf{V}^* = \text{Cov}(\mathbf{Y}^*) = \sigma_E^2\mathbf{I}_n + \sigma_A^2\mathbf{A}^*$,
and
$\mathbf{A}^* \in \mathbb{R}^{n\times n}$ is a block diagonal matrix with the $i$th block $\mathbf{1}_{n_{i}}\mathbf{1}_{n_{i} }^T$.
The $\sigma_A^2$ and $\sigma_E^2$ in \eqref{subbetahat} may also be replaced by their estimators trained from the subdata.
We will see that the accuracy of the estimators for $\sigma_A^2$ and $\sigma_E^2$ does not depend much on the subsampling strategies. Therefore, we will focus on selecting the subdata that allows the best estimation of $\boldsymbol{\beta}$.
From simple algebra, 
\begin{align*}
  \mathrm{E}(\hat{\boldsymbol{\beta}}^*) = \boldsymbol{\beta} \mbox{ and }
  \text{Var}(\hat{\boldsymbol{\beta}}^*) = (\mathbf{X}^{*T}\mathbf{V}^{*-1}\mathbf{X}^*)^{-1} = \mathbf{M}^{*-1},
\end{align*}
where
\begin{equation}\label{subm}
\mathbf{M}^* =\mathbf{X}^{*T}\mathbf{V}^{*-1}\mathbf{X}^*
\end{equation}
is the information matrix of the subdata. The optimal subdata $\mathbf{X}^*$ maximizes the information $\mathbf{M}^*$ or, in other words, minimizes Var$(\hat{\boldsymbol{\beta}}^*)$ in some manner, which can be obtained by minimizing an optimality function of $\mathbf{M}^{*-1}$. Denote $\psi$ as the optimality function. Finding the optimal subdata is to solve the following optimization problem:
\begin{align}\label{opt}
  \mathbf{X}^{*opt} &= \operatorname{arg} \underset{\mathbf{X}^{*}\subseteq\mathbf{X}}{\min}\ \psi(\mathbf{M}^{*-1})\nonumber\\
  &s.t. ~~ \mathbf{X}^{*} \mbox{ contains } n \mbox{ points. }
\end{align}
This is akin to the fundamental idea behind optimal experimental design \citep{Kiefer}. Popular options for $\psi$ include the determinant and trace, which correspond to the $D$- and $A$-optimality, respectively. Both of these two optimal criteria have specific statistical meanings. Specifically, $D$-optimal design minimizes the volume of the confidence ellipsoid centered at $\hat{\boldsymbol{\beta}}^*$  by maximizing the determinant $|\mathbf{M}^*|$, while $A$-optimal design minimizes the average variance of the components of $\hat{\boldsymbol{\beta}}^*$  by minimizing the trace $\mathrm{tr}(\mathbf{M}^{*-1})$.

The optimization problem in \eqref{opt} is not easy to solve. Exhaustive search for solving the problem requires $O(N^nn^2p)$ operations, which is infeasible for even moderate sizes of $\mathbf{X}$ and $\mathbf{X}^*$.
There are many types of algorithms for finding optimal designs and among them, exchange algorithms are among the most popular.
For the reasons argued in \cite{WangL}, these algorithms are cumbersome in solving the subsampling problem in \eqref{opt}.
To this end, we will initially derive theoretical results to establish the optimality of using an OA for the problem defined in \eqref{opt}. Following that, we will develop a computationally tractable subsampling approach called GOSS, which selects subdata approximating an OA. Consequently, instead of directly searching for the optimization in \eqref{opt}, GOSS efficiently utilizes an OA to approximate its solution.


\section{Optimality of OA for linear mixed model}\label{sec3}

An OA of strength $2$ on $s$ levels is a matrix with combinatorial orthogonality, that is, entries of the matrix come from a fixed finite set of $s$ levels, arranged in such a way that all ordered pairs of the levels appear equally often in every selection of two columns of the matrix.
For a comprehensive introduction to OA, see \cite{hedayat1999orthogonal}. In this paper, we consider $s = 2$, and denote the two levels by $-1$ and $1$. Here is an example of  $4 \times 3$ orthogonal array, where each of the ordered pairs $\{(-1,-1), (-1,1), (1,-1), (1,1)\}$ occurs once:
\begin{align*}
\begin{pmatrix}
-1 & -1 & -1 \\
-1 & 1 & 1 \\
1 & -1 & 1 \\
1 & 1 & -1
\end{pmatrix}.
\end{align*}
The combinatorial orthogonality of OA is actually a type of balance that ensures that all columns are considered fairly and rows distributed dissimilarly to cover as much different information as possible.
It has been shown that any OA with combinatorial orthogonality is simultaneously $D$- and $A$-optimal under a first-order linear model \citep{dey2009fractional}. These optimality properties of OA have been used in \cite{WangL} for subsampling problems under linear models.

Recall that in \eqref{opt}, for linear mixed model, the $D$-optimality criterion selects subdata that minimizes the determinant $|\mathbf{M}^{*-1}|$, that is, maximizes $|\mathbf{M}^*|$.
Notice that $\mathbf{V}^* = \text{diag}\{\mathbf{V}^*_{i}\}_{i=1}^{R}$, with $\mathbf{V}^*_{i} = \text{Cov}(\mathbf{Y}_{i}^{*})$ being the covariance matrix for the $i$th group, we thus can decompose $\mathbf{M}^*$ in \eqref{subm} by
\begin{align*}
    \mathbf{M}^*  = \sum_{i=1}^{R}\mathbf{X}_i^{*T}\mathbf{V}_i^{*-1}\mathbf{X}_i^* = \sum_{i=1}^{R}\mathbf{M}_i^*,
\end{align*}
where $\mathbf{M}_i^* = \mathbf{X}_i^{*T}\mathbf{V}_i^{*-1}\mathbf{X}_i^*$ is the information matrix for the $i$th group of the subdata. 
We first study the optimal $\mathbf{X}^*_i$ to maximize $|\mathbf{M}^*_{i}|$ when the number of points in $\mathbf{X}^*_i$ is given. To facilitate the presentation of the theoretical results below, without loss of generality, we assume that each covariate in $\mathbf{Z}_{i}$ has been scaled to $[-1, 1]$. 

\begin{lemma}\label{lem0}
For $i = 1, 2,\ldots, R$, 
let $n_i$ be the number of points in $\mathbf{X}^*_i$ and $\gamma_i = \sigma^2_E/(\sigma^2_E + n_{i} \sigma^2_A)$, then
\begin{align*}
|\mathbf{M}^*_{i}|\leqslant \gamma_{i} \left(\frac{n_{i} }{\sigma^{2}_E}\right)^p, 
\end{align*}
with equality if and only if $\mathbf{Z}^*_i$ forms a two-level $\mathrm{OA}$ with $n_i$ runs.
\end{lemma}

Lemma \ref{lem0} shows that given the number of points in $\mathbf{Z}^*_i$, it should form an OA to maximize $|\mathbf{M}^*_{i}|$.
To find the subdata that maximizes $|\mathbf{M}^*|$, we are concerned about two questions. First, following Lemma \ref{lem0}, does aggregating the OA subdata in each group maximize the overall information $|\mathbf{M}^*|$? Second, what are the optimal settings for $n_i$, $i=1,\ldots, R$? The following theorem, guiding our later algorithm, answers the two questions.

\begin{theorem}\label{th1.2}
For a subdata set $\mathbf{X}^*$ with $n$ points, $\mathbf{M}^*$ in \eqref{subm} satisfies that
\begin{align}
|\mathbf{M}^*| \leqslant \frac{n^{p-1}}{\sigma_E^{2p}} \left[\sum_{i=1}^{R}\gamma_in_i\right] \leqslant \frac{Rn^p}{\sigma_E^{2(p-1)}(R\sigma_E^{2} + n \sigma_A^{2})},\label{th1.2.1}
\end{align}
where $n_i$ is the number of points of the $i$th group in $\mathbf{X}_i^*$ and $\gamma_i = \sigma^2_E/(\sigma^2_E + n_{i} \sigma^2_A)$.
In addition, (i) the first equality in (\ref{th1.2.1}) holds when each $\mathbf{Z}^*_i$ forms a two-level $\mathrm{OA}$, and further,
                  (ii) the second equality holds if and only if the runsize of each $\mathrm{OA}$ selected from each group is equal, that is, $n_1 = \ldots = n_R$.
\end{theorem}

By Theorem \ref{th1.2}, the $D$-optimal subdata should have a group orthogonality, that is, equal-sized groups with each group forming an OA.
The following result shows that such group-orthogonal subdata is also $A$-optimal.

\begin{theorem}\label{co1}
For a subdata set $\mathbf{X}^*$ with $n$ points, $\mathbf{M}^*$ in \eqref{subm} satisfies that
\begin{align}
 \mathrm{tr}(\mathbf{M}^{*-1}) &\geqslant \sigma_E^2  \left(\frac{1}{\sum_{i=1}^{R}\gamma_in_i} + \frac{p-1}{n}\right)\label{co1.1} \\
 &\geqslant \frac{1}{n}\left (p \sigma_E^2 + \frac{n}{R} \sigma_A^2 \right),\label{co1.2}
\end{align}
where (i) the equality in (\ref{co1.1}) holds when each $\mathbf{Z}^*_i$ forms a two-level $\mathrm{OA}$, and
                  (ii) the equality in (\ref{co1.2}) holds if and only if the runsize of each $\mathrm{OA}$ selected from each group is equal, that is, $n_1 = \ldots = n_R$.
\end{theorem}

Theorems \ref{th1.2} and \ref{co1} suggest selecting the group-orthogonal subdata for fitting linear mixed models.
It is also worth noting that the optimal subdata is independent of ${\sigma}_A^2$ and ${\sigma}_E^2$. That is, we do not need to estimate ${\sigma}_A^2$ and ${\sigma}_E^2$ before subsampling, which further simplifies our calculation.
To this end, we propose the GOSS algorithm, which is specifically designed for hierarchical data and holds for any ${\sigma}_A^2$ and ${\sigma}_E^2$.

\section{Group-orthogonal subsampling}\label{sec3.2}
In this section, we propose the GOSS method. By the discussion in Section \ref{sec3}, the optimal subdata should have the same group size and form an OA in each group.
Recall that \cite{WangL} introduced the OSS algorithm to select subdata that best approximates an OA. Hence, GOSS can employ OSS to select the subdata from each group.
Specifically, we sequentially select data points from the $i$th group to minimize the discrepancy function:
\begin{align}\label{eqestimate1}
L\left(\mathbf{Z}_{i}^*\right)=\sum_{1 \leqslant j<j' \leqslant n_i}\left[(p-1)-\left\|\mathbf{z}_{ij}^*\right\|^{2} / 2-\left\|\mathbf{z}_{ij'}^*\right\|^{2} / 2+\delta\left(\mathbf{z}_{ij}^*, \mathbf{z}_{ij'}^*\right)\right]^{2},
\end{align}
where $$\delta\left(\mathbf{z}_{ij}^*, \mathbf{z}_{ij'}^*\right)=\sum_{k=2}^{p} \delta_{1}\left(x_{ijk}^*, x_{ij'k}^{*}\right),$$
and $\delta_{1}(x, y)$ is $1$ if both $x$ and $y$ have the same sign and $0$ otherwise. The function $L\left(\mathbf{Z}_{i}^*\right)$ measures the distance between $\mathbf{Z}_{i}^*$ and an OA. Therefore, the subdata for the $i$th group obtained by minimizing \eqref{eqestimate1} can well approximate an OA. The details of the OSS approach can be found in Section \ref{appC} of the Supplementary Materials.

Other than the orthogonality within each group, GOSS needs to make sure that the group size of the selected subdata are balanced. Therefore, for the desired subdata size $n$, we choose $m = n/R$ points from each group. After we have subdata from all groups, we aggregate all the subdata and obtain the GLS estimator for a linear mixed model.
Algorithm \ref{alg1} outlines the proposed GOSS algorithm.

\begin{algorithm}[ht!]
\caption{GOSS algorithm}\label{alg1}
\begin{algorithmic}
\REQUIRE
Full data $\mathbf{Z} = \left(\mathbf{Z}_1^T, \ldots, \mathbf{Z}_R^T\right)^T$, $\mathbf{Y} = \left(\mathbf{Y}_1^T, \ldots, \mathbf{Y}_R^T\right)^T$, subdata size $n$
\ENSURE  The subdata-based GLS estimator of $\breve{\boldsymbol{\beta}}^*$
\FOR{$i=1$ to $R$}
\STATE Let $m = n/R$. Use the OSS method
to minimize the discrepancy function in \eqref{eqestimate1} and select a subdata of size $m$ from group $i$, denoted as $\left\{\mathbf{Z}_{i}^*, \mathbf{Y}_{i}^*\right\}$
\ENDFOR
\STATE Aggregate the $R$ subdata sets as $\mathbf{Z}^* = \left(\mathbf{Z}_{1}^{*T}, \ldots, \mathbf{Z}_{R}^{*T}\right)^T$ and $\mathbf{Y}^* = \left(\mathbf{Y}_{1}^{*T}, \ldots, \mathbf{Y}_{R}^{*T}\right)^{T}$.\ Let $\hat{\sigma}_A^2$ and $\hat{\sigma}_E^2$ be consistent estimators of ${\sigma}_A^2$ and ${\sigma}_E^2$ based on the selected data $\mathbf{X}^* = \left(\mathbf{1}_{n}, \mathbf{Z}^*\right)$ and $\mathbf{Y}^*$. Estimate the coefficient $\boldsymbol{\beta}$ using
  \begin{equation}\label{eqestimate}
\breve{\boldsymbol{\beta}}^* = (\mathbf{X}^{*T}\hat{\mathbf{V}}^{*-1}\mathbf{X}^*)^{-1}\mathbf{X}^{*T}\hat{\mathbf{V}}^{*-1}\mathbf{Y}^*,
  \end{equation}
where $\hat{\mathbf{V}}^* = \hat{\sigma}_E^2\mathbf{I}_n + \hat{\sigma}_A^2\mathbf{A}^*$ and $\mathbf{A}^*$ is a block diagonal matrix with $R$ blocks of $\mathbf{1}_{m}\mathbf{1}_{m}^T$.
\end{algorithmic}
\end{algorithm}


\begin{remark}\label{rem1}
The restriction of Algorithm \ref{alg1} that $m=n/R$ is an integer is mostly for convenience. In
the case that $m=n/R$ is not an integer, we may use a combination of
$\lfloor m \rfloor$ and $\lceil m \rceil$ to keep the subdata size as $n$.

\end{remark}

\begin{remark}
We use the method of moments approach proposed by \cite{gao2017efficient} (refer to Section \ref{appD} in the Supplementary Materials) to estimate ${\sigma}_A^2$ and ${\sigma}_E^2$ in our numerical results in Sections \ref{sec4} and \ref{sec5}.
From Theorem 1 of \cite{gao2019estimation}, the moment method estimators based on GOSS subdata are consistent with variances
\begin{align*}
 Var(\hat{\sigma}^2_A) = O(R^{-1})\mbox{ and } \ Var(\hat{\sigma}^2_E) = O(m^{-1}).
\end{align*}
\end{remark}

\begin{remark}
The computation in Algorithm \ref{alg1} is mostly involved in OSS in each group, so the time complexity of Algorithm \ref{alg1} is $O(Np\ln m)$ \citep{WangL}. In addition, Algorithm \ref{alg1} is naturally suited for distributed and parallel computing. We can
simultaneously process each group of the full data, which will dramatically accelerate the subsampling process. 
\end{remark}

Compared to OSS, GOSS offers two main novel advantages. First, GOSS suggests that subsampling should be groupwise for hierarchical data, and the group size of the subdata should be the same. This is to ensure that the contribution of groups in the subdata are balanced. OSS, by contrast, directly subsamples the full data, resulting in unbalanced contributions from groups. Second, compared to OSS, which only ensures the combinatorial orthogonality of the entire subdata, GOSS further ensures the combinatorial orthogonality of the subdata in each group.
This groupwise orthogonality adds an additional layer of value to the subdata. As detailed in the theory presented in Section \ref{sec3}, it will significantly benefit the fitting of a linear mixed model.

Next, we discuss the asymptotic behavior of the slope estimator. Let $\boldsymbol{\beta} = (\beta_{1}, \boldsymbol{\beta}_{-1}^T)^T$, where $\beta_{1}$ is the intercept and $\boldsymbol{\beta}_{-1}$ the slope parameter. In practice, we are typically more interested in the estimation of $\boldsymbol{\beta}_{-1}$. Write the $\breve{\boldsymbol{\beta}}^*$  in \eqref{eqestimate} as $\breve{\boldsymbol{\beta}}^* = (\breve{\beta}_1^*, \breve{\boldsymbol{\beta}}_{-1}^{*T})^T$. We next study the asymptotic normality of $\breve{\boldsymbol{\beta}}_{-1}^{*}$ as an estimator of $\boldsymbol{\beta}_{-1}$.
Write the subdata design matrix $\mathbf{Z}^*_i$ from each group as $$\mathbf{Z}^*_i = {\mathbf{L}}_i^* + {\mathbf{D}}_i^*,$$ where ${\mathbf{L}}_i^*$ is a two-level OA, and ${\mathbf{D}}_i^*$ is the difference between $\mathbf{Z}^*_i$ and $\mathbf{L}_i^*$. Let $\mathbf{D}^*= (\mathbf{D}_{1}^{*T},\ldots,\mathbf{D}_{R}^{*T})^T$ and $||{\mathbf{D}}^*||_{\infty}$ be the entrywise max norm, i.e., the maximum absolute value of the entries in ${\mathbf{D}}^*$. We have the following theorem.

\begin{theorem}\label{th4}
For a fixed number of groups $R$, suppose that the maximum norm of $\mathbf{D}^*$ is $||{\mathbf{D}}^*||_{\infty} = o(1)$ as $n = Rm \rightarrow \infty$, $\mathrm{E}|e_{ij}^3| < \infty$, and $\hat{\sigma}_A^2$ and $\hat{\sigma}_E^2$ are consistent estimators of ${\sigma}_A^2$ and ${\sigma}_E^2$ respectively. For the estimator of the slope parameter in \eqref{eqestimate}, $\breve{\boldsymbol{\beta}}_{-1}^{*}$, we have 
\begin{align*}
  \sqrt{n}\left(\breve{\boldsymbol{\beta}}_{-1}^{*} - \boldsymbol{\beta}_{-1}\right) \stackrel{d}{\longrightarrow}  N(\mathbf{0}, \sigma_E^{2}\mathbf{I}_{p-1}), ~~~as \ n \rightarrow \infty,
\end{align*}
where $``\stackrel{d}{\longrightarrow}"$ denotes convergence in distribution.
\end{theorem}

Theorem \ref{th4} indicates that the slope estimator based on a GOSS subdata is asymptotically normal with a covariance matrix $\sigma_E^{2}\mathbf{I}_{p-1}$ and an average variance $\sigma_E^{2}$, which is the smallest possible average variance for an estimator of $\boldsymbol{\beta}_{-1}$. Because the subdata size $n$ is typically finite, the smaller asymptotic variance guarantees that the estimator based on a GOSS subdata is more accurate than other subdata.

\section{Simulation studies}\label{sec4}
In this section, we evaluate the performance of GOSS with simulation studies.
Let the number of groups $R=20$. The first $10$ groups have the same data size, and the last $10$ groups have the same data size, that is, $C_1=\cdots=C_{10}$ and $C_{11}=\cdots=C_{20}$.
Four cases are considered to generate the design matrix $\mathbf{Z}=(\mathbf{z}_{ij,k})$ of the full data for $j=1,\ldots, C_i$, $i=1,\ldots, 20$, and $k=2, \ldots, p$. Cases 1 and 2 consider homogeneous data, where data in all groups are from an identical distribution.
Cases 3 and 4 consider heterogeneous data with different group means, simulating heterogeneity among the groups. Specifically, we consider the following settings:

\begin{description}
\item[Case 1.]  The covariates $\mathbf{z}_{ij}$'s are independent and follow a multivariate uniform distribution: $\mathbf{z}_{ij,k}\sim U[-1,1],$ $k =2, \ldots, p.$
\item[Case 2.] The covariates $\mathbf{z}_{ij}$'s follow a multivariate normal distribution: $\mathbf{z}_{ij}\sim N(\mathbf{0},\mathbf{\Sigma})$, with
$$\mathbf{\Sigma} = \left(0.5^{I(k\neq k')}\right), k, k' =2, \ldots, p.$$
\item[Case 3.] The covariates $\mathbf{z}_{ij}$'s follow a uniform distribution: $\mathbf{z}_{ij,k}\sim U[\theta_{i1}, \theta_{i2}]$, where $U[\theta_{i1}, \theta_{i2}]$ is a shift of $U[-1,1]$ such that the centers of groups vary within $\{-0.5, -0.45, \ldots,0.45\}$. Thus, we set $\theta_{i1}=-1+(i-11)/20$ and $\theta_{i2}=1+(i-11)/20$.
\item[Case 4.] The covariates $\mathbf{z}_{ij}$'s follow a multivariate normal distribution: $\mathbf{z}_{ij}\sim N(\mu_i\mathbf{1},\mathbf{\Sigma})$, with $\mu_i$ varying within $\{-2, -1.8, \ldots, 1.8\}$.
\end{description}
The response data are generated from the linear mixed model (\ref{1}) with the true value of $\boldsymbol{\beta}$ being a $51 \times 1$ vector of unity which includes an intercept and fifty slope parameters, so $p=51$. The error term is generated from $e_{ij}\sim N(0,9)$. We consider two settings of the random effect, namely, $a_{i}\sim N(0, 0.5)$ and $a_{i}\sim t(3)$, to illustrate the impact of the distribution and variance of the random effect. Here $a_{i}\sim N(0, 0.5)$ simulates smaller random effects and thus lower correlations between responses within groups, while $a_{i}\sim t(3)$ simulates larger random effects and higher correlations within groups.

\subsection{Comparison of performance}
The simulation is repeated for $B = 200$ times. We compare the following different subsampling methods:  UNIF (simple random subsampling with uniform weights), LEV (leveraging subsampling), IBOSS, OSS, GUNIF (Group-UNIF), GLEV (Group-LEV), GIBOSS (Group-IBOSS), and GOSS.
The GUNIF, GLEV, and GIBOSS methods select the same number of data from each group using the UNIF, LEV, and IBOSS methods respectively. We compare these three methods with the GOSS algorithm to demonstrate that the optimality of GOSS is not merely attributed to the balance of subdata sizes among groups, but also to the orthogonality of the subdata within each group.
For each subsampling method, we consider the empirical MSE of the slope parameters:
\begin{equation}\label{mse1}
\mathrm{MSE} = B^{-1}\sum_{b =1}^{B} ||\breve{\boldsymbol{\beta}}_{-1}^{*(b)} - \boldsymbol{\beta}_{-1}||^2,
\end{equation}
where $\breve{\boldsymbol{\beta}}_{-1}^{*(b)}$ is the GLS estimator of $\boldsymbol{\beta}_{-1}$ based on subdata in the $b$th repetition. 

We first consider the setting of $C_1=\cdots=C_{10}=5\times 10^3$ and $C_{11}=\cdots=C_{20}= 2C_1$, resulting in a fixed full data size of $N = 1.5\times 10^5$.
Since $\sigma^2_A$ and $\sigma^2_E$ are unknown in practice, we estimate them based on subdata using the moment method proposed by \cite{gao2017efficient} and plug them into the estimator $\breve{\boldsymbol{\beta}}_{-1}^{*(b)}$.
Figure \ref{supp::fig1} in Supplementary Materials shows the $\text{log}_{10}\text{(MSE)}$ of $\hat{\sigma}^2_A$ and $\hat{\sigma}^2_E$ with respect to subdata sizes $n = 10^3, 2\times 10^3, 3\times 10^3,$ and $4\times 10^3$ when $a_{i}\sim N(0, 0.5)$. We observe that all the subdata tend to provide reliable estimates for ${\sigma}^2_A$ and ${\sigma}^2_E$, except for OSS in Case 3 when the subdata size is small ($n=1000$).

With $\hat{\sigma}^2_A$ and $\hat{\sigma}^2_E$, Figure \ref{fig4} plots the $\log_{10}$(MSE) of the plug-in estimator $\breve{\boldsymbol{\beta}}_{-1}^{*(b)}$ with respect to $n$.
For Cases 1 and 2, grouped methods perform similarly to their counterparts because groups are identically distributed, and GOSS and OSS outperform other methods due to the orthogonality of the subdata.
For Cases 3 and 4, however, the performance of GOSS dominates all other methods for every subdata size $n$, although all methods decrease at the same rate.
It should be noted that GUNIF and GIBOSS do not outperform their counterparts, indicating that the advantages of the GOSS method go beyond the balancing of group sizes, and within-group orthogonality is crucial in determining its superiority.
Moreover, the fact that the GOSS method outperforms other methods in both the upper and lower panels of Figure \ref{fig4} demonstrates that GOSS is powerful regardless of the size of random effects.



We also consider the performance of GOSS for different full data sizes and show the result in Figure \ref{fig3}.
We consider $C_1=\cdots=C_{10} \in \{10^3, 5\times10^3, 2.5\times 10^4, 1.25\times10^5\}$ and $C_{11}=\cdots=C_{20}=2C_1$, which results in the full data size $N \in \{3\times 10^{4}, 1.5\times 10^{5}, 7.5\times 10^{5}, 3.75\times 10^{6}\}$.
The subdata size is fixed at $n = 4\times 10^3$.
As evidenced by Figure \ref{fig3}, for Cases 1 and 2, grouped methods perform similarly to their counterparts, and both GOSS and OSS exhibit outstanding performance and fast decreasing MSEs as $N$ increases, meaning that they can both extract more information from the full data as the size of the full data increases.
For Case 3, OSS fails to extract more information as $N$ increases because of the heterogeneity of the full data, but GOSS keeps its fast decreasing trend and outperforms all other methods significantly.
For Case 4, the GOSS method retains its remarkable superiority, even though the IBOSS and GIBOSS also exhibit a slow decreasing trend.


We further examine the performance of GOSS when there is an extreme imbalance among group sizes in full data.
To this end, we change the setting of $C_i$ to $C_1=\cdots=C_{10}=5\times10^3$ and $C_{11}=\cdots=C_{20}= 10C_1=5\times10^4$.
Figure \ref{supp::figC} in Supplementary Materials
plots $\text{log}_{10}\text{(MSE)}$ for $\hat{\sigma}^2_A$ and $\hat{\sigma}^2_E$ with respect to the subdata size $n$, and Figure \ref{fig5} shows the $\text{log}_{10}\text{(MSE)}$ for $\breve{\boldsymbol{\beta}}_{-1}^{*(b)}$ versus  $n$.
The GOSS still outperforms all other methods for Cases 3 and 4 because of its balance among groups and within-group orthogonality, which still provides more information even though the group sizes of the full data are extremely unbalanced.





To see the performance of GOSS when the full data size grows and is extremely imbalanced, we further consider $C_1=\cdots=C_{10} \in \{10^3, 5\times10^3, 2.5\times 10^4\}$ and $C_{11}=\cdots=C_{20}=10C_1$, with the full data size $N \in \{1.1\times 10^{5}, 5.5\times 10^{5}, 2.75\times 10^{6}\}$. The subdata size is again fixed at $n = 4\times10^3$.
According to Figure \ref{fignew1}, all subsampling methods behave similarly as in Figure \ref{fig3}. One point to note is that for Case 2, 
the grouped methods appear to be slightly inferior to their counterparts, mainly because of the homogeneous and overlapping information in all groups of the full data. In this case, drawing the same amount of information from each group can result in missing more important information in bigger groups.
For Cases 3 and 4,
the superiority of GOSS is attributed to the balance of heterogeneous groups, which contain information from different aspects. The balance among these groups enables more accurate modeling and parameter estimation, resulting in a fast downward trend and improved performance.

We have also conducted simulations to evaluate the performance of subsampling methods in estimating the intercept and predicting the response over the full data. Possible model misspecification has also been considered. Due to page limitations, the results are deferred to Section \ref{appB} of the Supplementary Materials. 



\subsection{Computing time}
Table \ref{tab1} reports the computation times (including the selection of subdata and the computation of estimators of $\boldsymbol{\beta}$, in seconds) under the setting of $C_1=\cdots=C_{10} = 5\times10^3$, $C_{11}=\cdots=C_{20}=2C_1$, $p = 6, 51,$ and $101$, and $n=10^3$. Covariates are generated as in Case 3 and the random effect $a_{i}\sim N(0, 0.5)$.
The times shown in Table \ref{tab1} are the mean CPU times of $200$ repetitions. All computations are carried out on a laptop running Windows 10 21H2 with a 3.00GHz Intel Core i7 processor and 16GB memory. As indicated in Table \ref{tab1}, the grouped methods are more time-efficient than the ungrouped method. UNIF and GUNIF require the least computation time as expected. The GOSS is faster than LEV, OSS, and IBOSS and is comparable to GLEV and GIBOSS.
Table \ref{tab2} reports the computation times for different full data sizes $N$ with a fixed dimension $p = 51$ and a fixed subdata size $n=1000$. 
The GOSS is faster than LEV, OSS, and GIBOSS and is comparable to IBOSS and GLEV for all full data sizes.

\begin{table}[!h]
\centering
\caption{The CPU times (in seconds) of subsampling methods with $n = 10^3$.}
\setlength\tabcolsep{2.5mm}{
\begin{tabular}{lllllllll}
\hline
Method & UNIF & LEV & IBOSS & OSS  & GUNIF & GLEV & GIBOSS & GOSS \\ \hline
$p$ = 6 & 0.2240 & 0.2297 & 0.2001 & 0.2602 & 0.0883 & 0.1431 & 0.1313 & 0.1373\\
$p$ = 51 & 0.6006 & 1.2980 & 1.5579 & 1.8271 & 0.3936 & 0.8973 & 0.9745 & 0.8799\\
$p$ = 101 & 0.9349 & 3.6877 & 2.9458 & 3.6636 & 0.7431 & 1.7489 & 1.8859 & 1.7723\\\hline
\end{tabular}}
\label{tab1}
\end{table}

\begin{table}[!h]
\centering
\caption{The CPU times (in seconds) of subsampling methods with $p = 51$.}
\setlength\tabcolsep{1.8mm}{
\begin{tabular}{lllllllll}
\hline
Method & UNIF & LEV & IBOSS & OSS  & GUNIF & GLEV & GIBOSS & GOSS  \\ \hline
$N = 3\times10^4$ & 0.1347 & 0.1925 & 0.2984 & 0.5159 & 0.0981 & 0.1868 & 0.1867 & 0.1837\\
$N = 7.5\times10^5$ & 1.2927 & 2.7587 & 1.8938 & 2.8484 & 0.6611 & 2.0032 & 2.7679 & 1.9937 \\
$N = 3.75\times10^6$ & 6.3441 & 14.3277 &  8.8961 & 11.3674 &  3.0972 &  9.3464 & 17.8353 &  9.3434\\\hline
\end{tabular}}
\label{tab2}
\end{table}


\section{Real data analysis-Accelerometer dataset}\label{sec5}
We analyze the accelerometer dataset to evaluate the performance of the GOSS approach.
The data records the vibration of the cooler fan with weights on its blades, which allows us to infer when the motor failed.
To generate different vibration scenarios, the experimenters set $17$ different cooler fan speeds ranging from $20 \%$ to $100 \%$ of the maximum fan speed at  $5 \%$ intervals.
Vibrations were measured by accelerometers at a frequency of $20$ milliseconds, with vibration measurements taking 1 minute at each speed and generating $3,000$ recordings at each frequency. Thus, a total of $N = 153,000$ vibration records were collected.
Further details about the data can be found at \cite{scalabrini2019prediction}.
At each speed, the accelerometer measures $9000$ observations of vibration on $x$, $y$, and $z$ axes.
We grouped the data according to the $17$ different cooler fan speeds. Thus, the number of groups is $R=17$.
For each speed, the vibration on the $z$ axis varies with the vibration on the $x$ and $y$ axes.
We take the $x$ and $y$ axes as independent variables and the $z$ axis as the response variable to assess the impact of $x$ and $y$ axes vibrations on the $z$ axis. 
We consider the model
         \begin{equation}\label{RD}
z_{ij} = \beta_0 + \beta_1 x_{ij} + \beta_2 y_{ij} + a_i + e_{ij}, \  i = 1,\cdots,17, j=1,\cdots,9000,
        \end{equation}
where $a_i$ denotes the random effect of the cooler fan speed, and $e_{ij}$ is the random error of the response at the same speed. 

\begin{figure}[htpt]
\centering
\includegraphics[height=5cm]{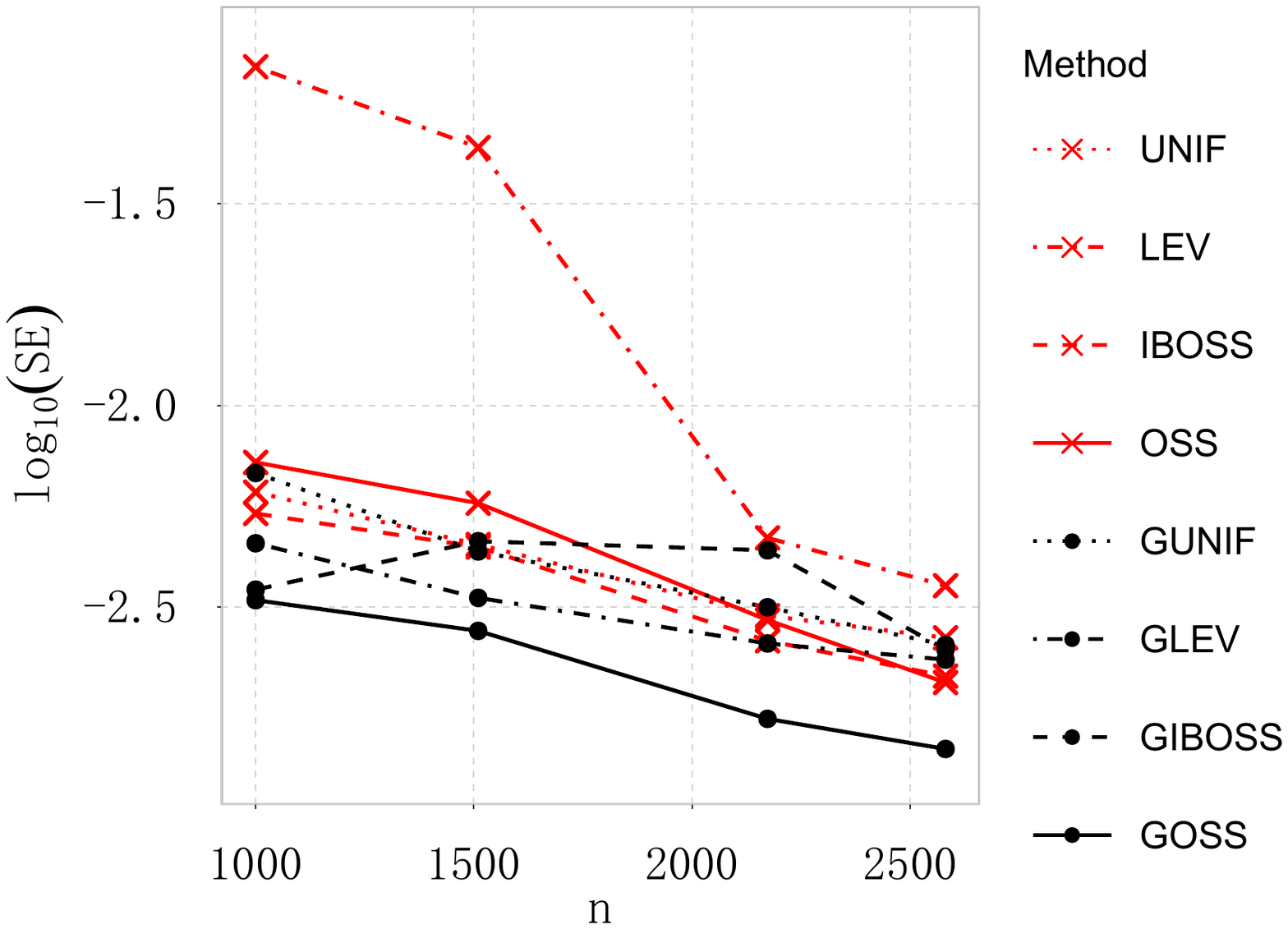}
\caption{The $\log_{10}$(SE) of $\breve{\boldsymbol{\beta}}^*_{-1}$ with different subdata sizes for the accelerometer dataset.}
	\label{fig6}
\end{figure}

\begin{figure}[htbp]
\centering
\includegraphics[width=18cm]{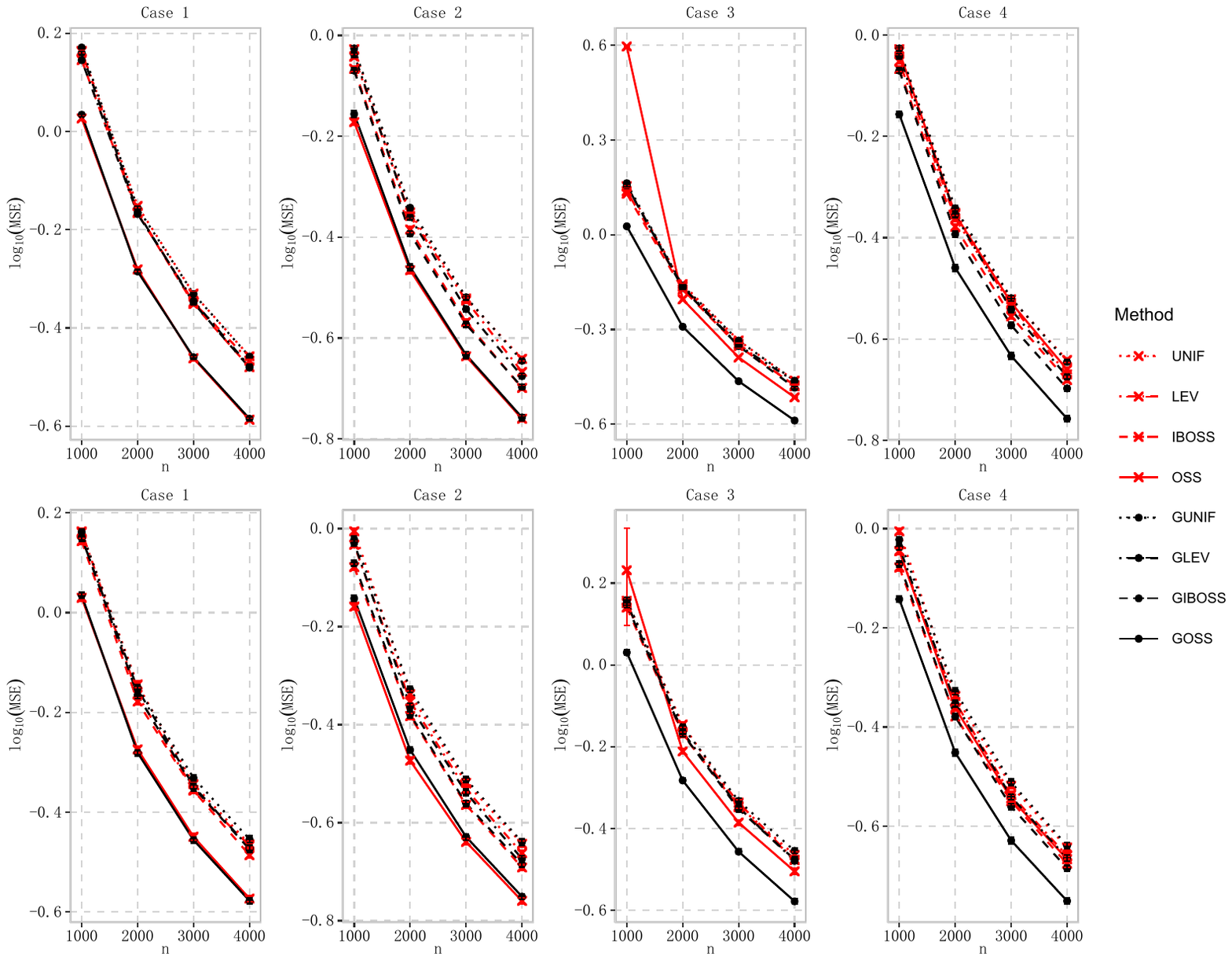}
	\caption{The $\text{log}_{10}$(MSE) of the estimated slope parameters for different subdata size $n$. The upper panels are for $a_{i}\sim N(0, 0.5)$ and the lower panels for $a_{i}\sim t(3)$. The full data size is $N=1.5\times 10^5$. The bars represent standard errors obtained from $200$ replicates. Some bars are very narrow, so they seem to be invisible.}
	\label{fig4}
\end{figure}

\begin{figure}[htbp]
\centering
\includegraphics[width=18cm]{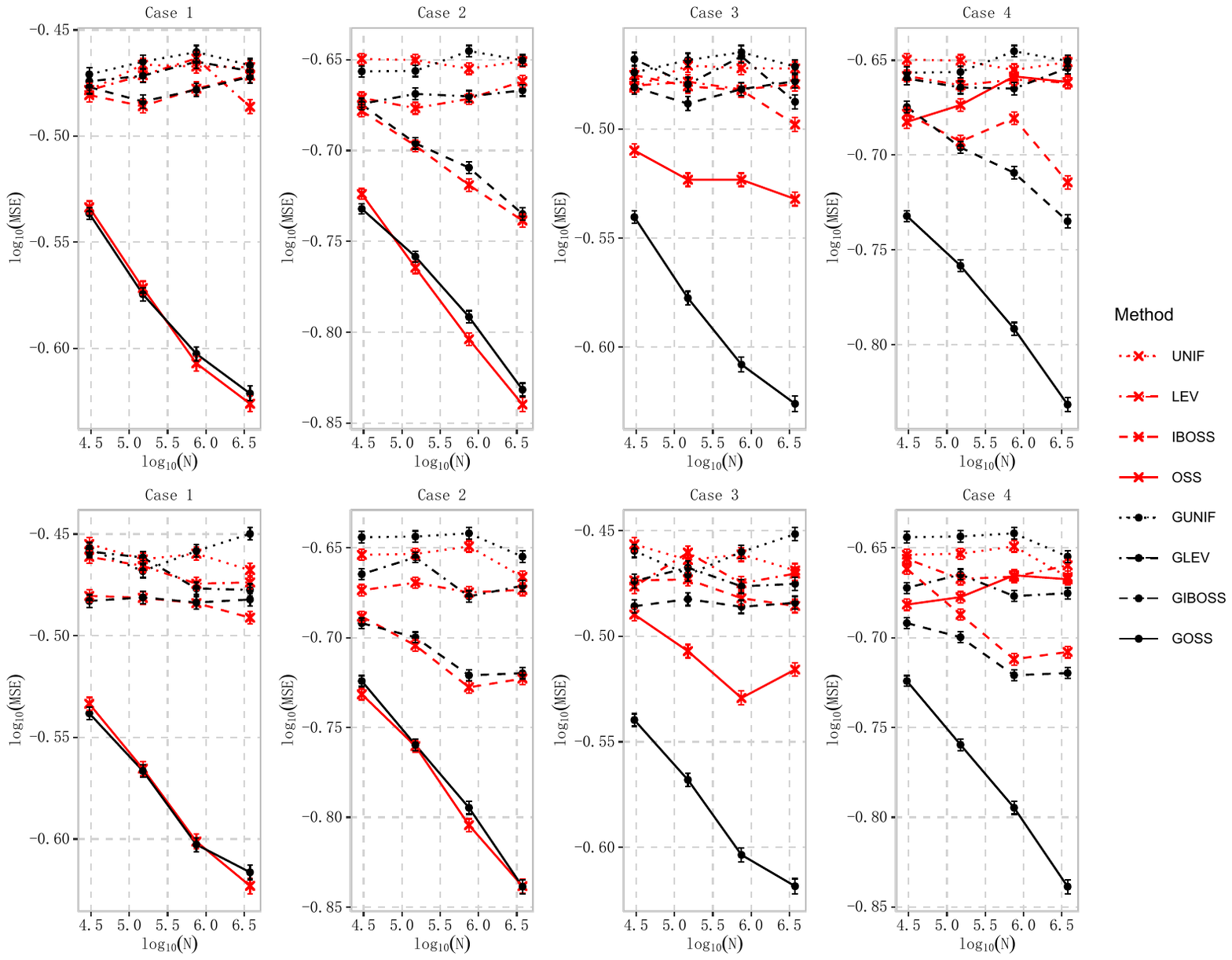}
\caption{The $\text{log}_{10}$(MSE) of the estimated slope parameters for different full data size $N$. The subdata size is fixed at $n = 4\times 10^3$. The upper panels are for $a_{i}\sim N(0, 0.5)$, and the lower panels for $a_{i}\sim t(3)$. The bars represent standard errors obtained from $200$ replicates.}
\label{fig3}
\end{figure}

\begin{figure}[htbp]
\centering
\includegraphics[width=18cm]{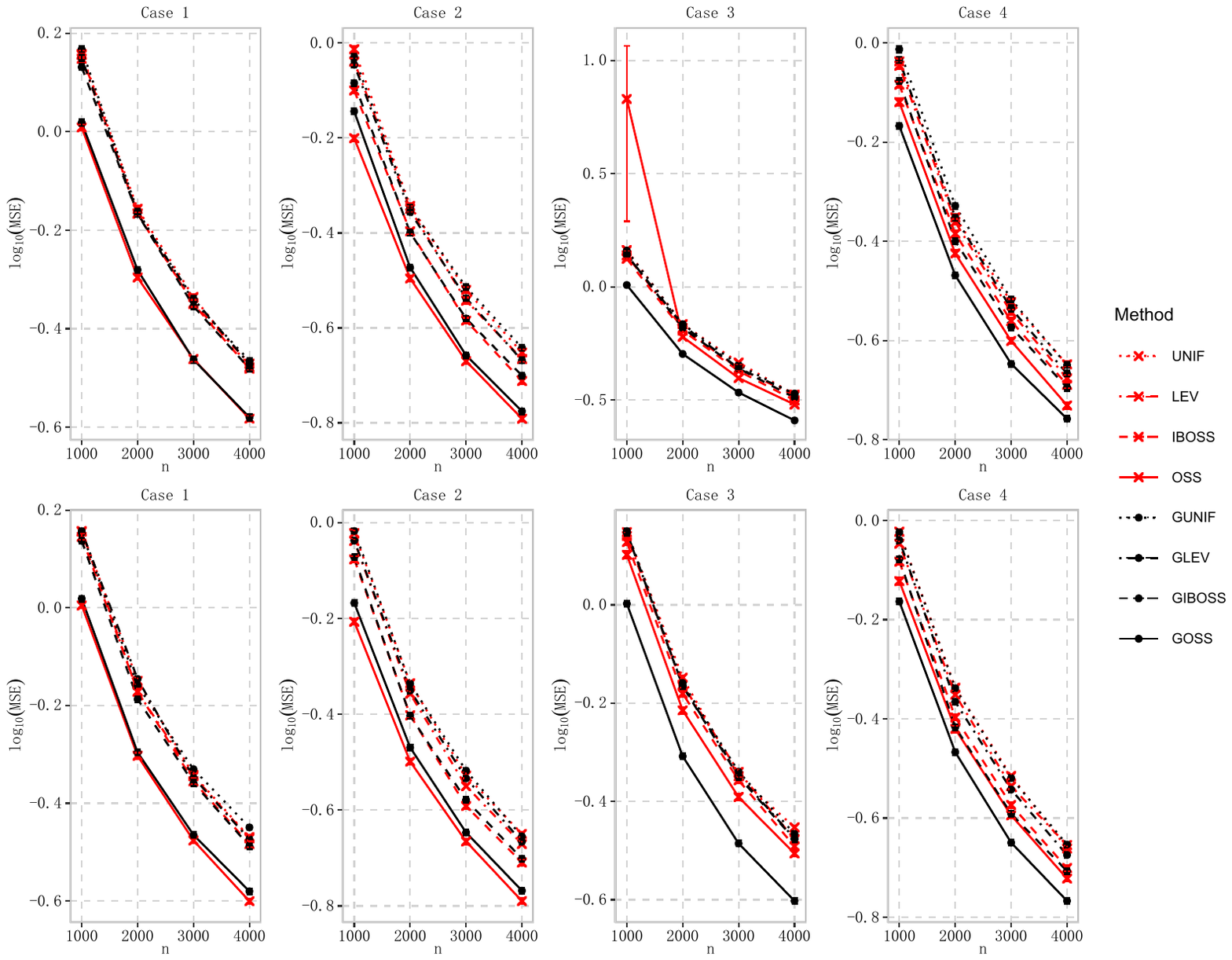}
	\caption{The $\text{log}_{10}$(MSE) of the estimated slope parameters for different subdata size $n$. The upper panels are for $a_{i}\sim N(0, 0.5)$ and the lower panels for $a_{i}\sim t(3)$. The full data size is $N=5.5\times 10^5$\reva{.} The bars represent standard errors obtained from $200$ replicates. Some bars are very narrow and may be invisible.}
	\label{fig5}
\end{figure}

\begin{figure}[htbp]
\centering
\includegraphics[width=18cm]{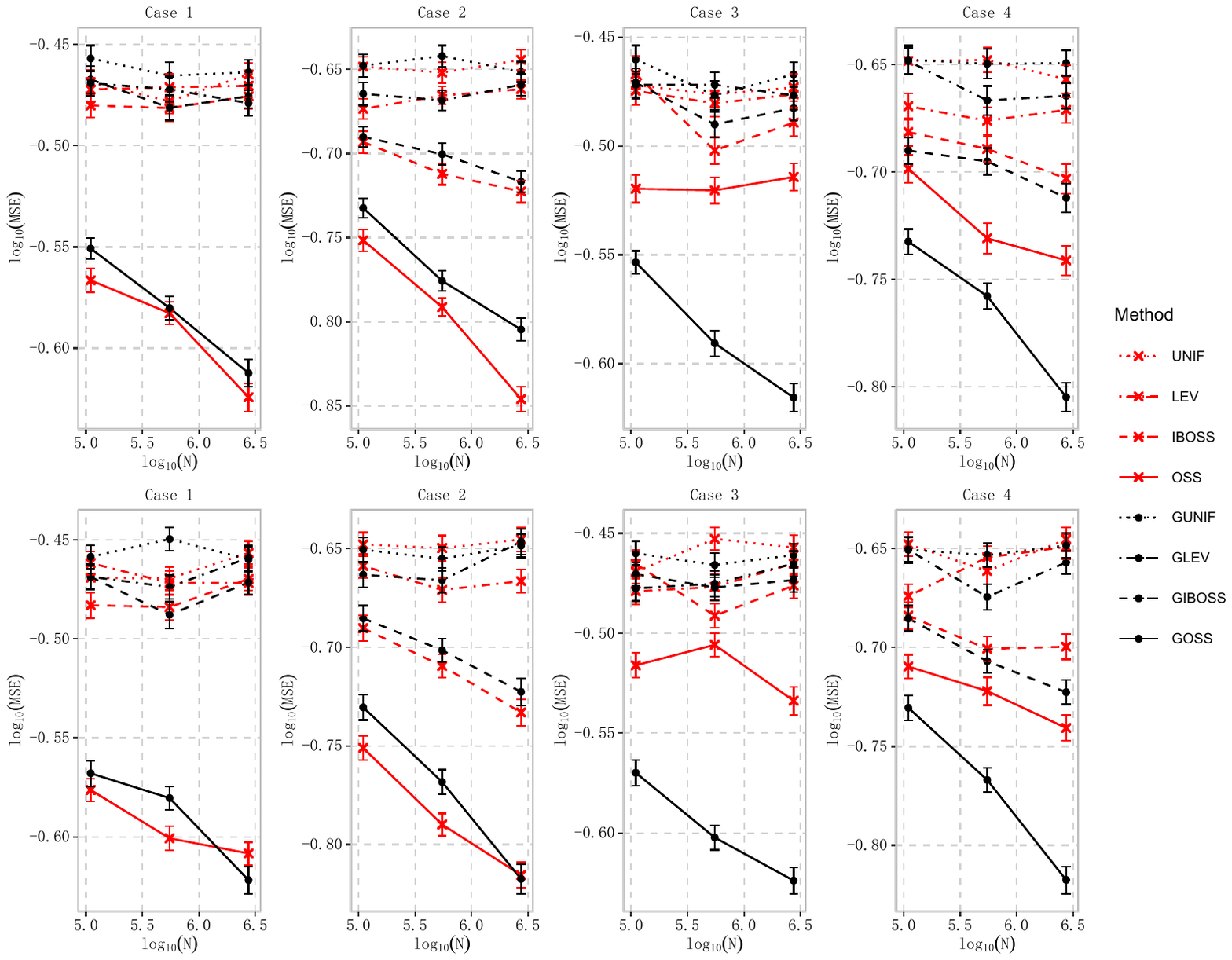}
\caption{The $\text{log}_{10}$(MSE) of the estimated slope parameters for different full data sizes $N$, when there is an extreme imbalance in the data sizes among groups. The subdata size is fixed at $n = 4\times 10^3$. The upper panels are for $a_{i}\sim N(0, 0.5)$, and the lower panels for $a_{i}\sim t(3)$. The bars represent standard errors obtained from $200$ replicates.}
\label{fignew1}
\end{figure}

We consider subdata sizes $n=1000, 1510, 2173$, and $2581$ and assess subsampling methods by examining the difference between the estimator derived from subdata and the estimator obtained from the full data. That is, we consider the squared error (SE)
$$SE = ||\breve{\boldsymbol{\beta}}^{*}_{-1} - \hat{\boldsymbol{\beta}}_{-1}||^2,$$
where $\hat{\boldsymbol{\beta}}_{-1}$ is the GLS estimator of the slope parameter $\boldsymbol{\beta}_{-1} = (\beta_1, \beta_2)^T$ based on the full data, and $\breve{\boldsymbol{\beta}}^{*}_{-1}$ is the estimator from subdata. For the methods UNIF, LEV, GUNIF, and GLEV, we repeat them 200 times due to their randomness and calculate the average SE. OSS, IBOSS, GOSS, and GIBOSS are deterministic methods and are executed only once. Figure \ref{fig6} plots the SE for different subsampling methods. It is clear that GOSS outperforms all other methods for all subdata sizes in terms of minimizing the SE. Further, the SE for GOSS decreases fast as the subdata size increases, which suggests that GOSS allows a better estimation of the impact of $x$ and $y$ axes vibration on the vibration of the $z$ axis.

Table \ref{tab3} shows the CPU times (average over the 200 repetitions) of different subsampling methods for the accelerometer data with $n = 1000$.
The comparison in Table \ref{tab3} is consistent with that in Table \ref{tab1}, which again shows that GOSS is faster than OSS and IBOSS and is comparable to GIBOSS.

\begin{table}[htpt]
\centering
\caption{The real data CPU times (in seconds) of subsampling methods with $n = 1000$.}
\setlength\tabcolsep{2.2mm}{
\begin{tabular}{llllllllll}
\hline
Method & UNIF & LEV  & IBOSS & OSS & GUNIF & GLEV & GIBOSS  & GOSS  & Full\\ \hline
Time  & 0.1400 & 0.2133 & 0.4142 & 0.5555 & 0.1491 & 0.2844 & 0.3319 & 0.2997 & 426 \\ \hline
\end{tabular}}
\label{tab3}
\end{table}

\section{Concluding remarks}\label{sec6}
In this paper, we present a novel subsampling method called GOSS, which is designed for selecting subdata from large datasets with a hierarchical structure. GOSS achieves data size balance among groups and combinatorial orthogonality within each group, ensuring that the selected subdata is $D$- and $A$-optimal for the GLS estimator of a linear mixed model. Extensive simulations and a real-world application demonstrate that GOSS outperforms existing methods in minimizing the MSE of the estimator for the slope parameter, especially in cases where data groups are heterogeneous. Theoretical results establish that the estimator obtained from the GOSS subdata has the minimum variance among all possible subdata, as evidenced by its asymptotic distribution. Additionally, GOSS is faster than competing methods, making it a highly efficient option for accelerating the analysis of big data using a linear mixed model.

Particular aspects associated with this research require more extensive and thorough studies.
First, GOSS is developed for scenarios where the full dataset has a fixed number of groups, with the sample size in each group tending toward infinity. However, in real-world applications, we may encounter situations where the number of groups tends toward infinity, while the sample size of each group remains limited. Subsampling methods that can handle this scenario require further study.
Second, we have only considered a constant within-group variance for convenience, but it is also common to have varying within-group variances, and addressing this issue is of pressing concern for future research.
Third, the data within each group may be sparse or incomplete due to missing values. Investigating suitable subsampling methods to handle sparse and incomplete data is another valuable avenue for exploration.

\section*{Supplementary Materials}

\begin{description}

\item[Online Appendix:]
provides proofs of the theoretical results in the main paper, additional numerical results, the OSS algorithm, and an estimation method for $\sigma_A^2$ and $\sigma_E^2$.

\item[Code and Data Zip File:] provides R code and data to replicate our results and apply the method to other dataset.

\end{description}

\section*{Acknowledgements}
This work was supported by the NSFC Grant (Nos.~11971098, 11471069) and the National Key Research and Development Program of China (Nos. 2020YFA0714102, 2022YFA1003701).


\clearpage

\begin{center}
{\large\bf Supplementary Materials for\\
``Group-Orthogonal Subsampling for Hierarchical Data Based on Linear Mixed Models"}
\end{center}

This document provides proof of theoretical results in the main paper, additional numerical results, the OSS algorithm, and an estimation method for $\sigma_A^2$ and $\sigma_E^2$.

\begin{appendix}
\setcounter{figure}{0}
\renewcommand{\thefigure}{S\arabic{figure}}
\setcounter{equation}{0}
\renewcommand{\theequation}{S\arabic{equation}}
\setcounter{algorithm}{0}
\renewcommand{\thealgorithm}{S\arabic{algorithm}}
\setcounter{remark}{0}
\renewcommand{\theremark}{S\arabic{remark}}
\renewcommand{\theequation}{S\arabic{equation}}
\setcounter{lemma}{0}
\renewcommand{\thelemma}{S\arabic{lemma}}

\section{Technical proofs}\label{appA}
Before presenting the proof of Lemma \ref{lem0}, we first state two essential lemmas.

\begin{lemma}\label{supp::lem1}
 Let $\mathbf{T} \in \mathbb{R}^{u \times v}$ be a matrix with elements contained in $[-1, 1]$. Then
\begin{align*}
|\mathbf{T}^T\mathbf{T}| \leqslant u^v,
\end{align*}
where the equality holds if and only if $\mathbf{T}$ forms a two-level $\mathrm{OA}$ with $u$ runs.
\end{lemma}
\textbf{Proof}
Denote $\mathbf{T} = (\mathbf{T}_1, \ldots, \mathbf{T}_v)$, where $\mathbf{T}_i$ is the $i$th column of $\mathbf{T}$. We have
\begin{align}
|\mathbf{T}^T\mathbf{T}| = \prod_{k=1}^{v} \lambda_k \leqslant \left(\frac{1}{v}\sum_{k=1}^{v} \lambda_k\right)^v \leqslant \left(\frac{1}{v}tr\left(\mathbf{T}^T\mathbf{T}\right)\right)^v = \left(\frac{1}{v}\sum_{i=1}^{v} ||\mathbf{T}_i||^2\right)^v \leqslant u^v, \label{supp::le1.1}
\end{align}
where $\lambda_k, k = 1, 2,\cdots,v$ are eigenvalues of $\mathbf{T}^T\mathbf{T}$, $||\cdot||$ represents the Euclidean norm. The last inequality in (\ref{supp::le1.1}) comes from the fact that the elements of $\mathbf{T}$ contained in $[-1, 1]$. Then the inequalities in (\ref{supp::le1.1}) become equalities if and only if $\lambda_1 = \lambda_2 = \ldots =\lambda_v = u$, at this time $\mathbf{T}^T\mathbf{T} = u \mathbf{I}_v$.

Thus, $|\mathbf{T}^T\mathbf{T}| =  u^v$ if and only if $\mathbf{T}$ forms a two-level OA with $u$ runs.  This completes the proof.

\begin{lemma}\label{supp::lem2}
 Let $\tilde{\mathbf{T}} = (\mathbf{1}_u, \mathbf{T})$, where $\mathbf{T}$ is defined in Lemma \ref{supp::lem1}. $c$ is a constant contained in $(0, 1)$. Then
\begin{align*}
|\tilde{\mathbf{T}}^{T}\tilde{\mathbf{T}} - cu^{-1}\tilde{\mathbf{T}}^{T}\mathbf{1}_u\mathbf{1}_u^{T}\tilde{\mathbf{T}}| \leqslant (1-c)u^{v+1},
\end{align*}
where the equality holds if and only if $\mathbf{T}$ forms a two-level $\mathrm{OA}$ with $u$ runs.
\end{lemma}
\textbf{Proof} Note that $\tilde{\mathbf{T}} = (\mathbf{1}_u, \mathbf{T}_1, \ldots, \mathbf{T}_v)$. Let $T_{i\cdot}$ be the column sum of $\mathbf{T}_i$. After some simple calculations, $\tilde{\mathbf{T}}^{T}\tilde{\mathbf{T}} - cu^{-1}\tilde{\mathbf{T}}^{T}\mathbf{1}_u\mathbf{1}_u^{T}\tilde{\mathbf{T}}$ can be expressed as follows,
\begin{align*}
\tilde{\mathbf{T}}^{T}\tilde{\mathbf{T}} - cu^{-1}\tilde{\mathbf{T}}^{T}\mathbf{1}_u\mathbf{1}_u^{T}\tilde{\mathbf{T}} =
\begin{pmatrix}
  (1-c)u & (1-c)T_{1\cdot} & \cdots & (1-c)T_{v\cdot}\\
  (1-c)T_{1\cdot}& \mathbf{T}_1^T\mathbf{T}_1 - cu^{-1}T_{1\cdot}^2 & \cdots & \mathbf{T}_1^T\mathbf{T}_v - cu^{-1}T_{1\cdot}T_{v\cdot}\\
  \vdots & \vdots &  & \vdots \\
  (1-c)T_{v\cdot}& \mathbf{T}_v^T\mathbf{T}_1 - cu^{-1}T_{v\cdot}T_{1\cdot} & \cdots & \mathbf{T}_v^T\mathbf{T}_v - cu^{-1}T_{v\cdot}^2
\end{pmatrix}.
\end{align*}

Thus, the determinant of $\tilde{\mathbf{T}}^{T}\tilde{\mathbf{T}} - cu^{-1}\tilde{\mathbf{T}}^{T}\mathbf{1}_u\mathbf{1}_u^{T}\tilde{\mathbf{T}}$ can be expressed as follows,
\begin{align}
|\tilde{\mathbf{T}}^{T}\tilde{\mathbf{T}} - cu^{-1}\tilde{\mathbf{T}}^{T}\mathbf{1}_u\mathbf{1}_u^{T}\tilde{\mathbf{T}}|
&=
\begin{vmatrix}
  (1-c)u & (1-c)T_{1\cdot} & \cdots & (1-c)T_{v\cdot}\\
  0& \mathbf{T}_1^T\mathbf{T}_1 - u^{-1}T_{1\cdot}^2 & \cdots & \mathbf{T}_1^T\mathbf{T}_v - u^{-1}T_{1\cdot}T_{v\cdot}\\
  \vdots & \vdots &  & \vdots \\
  0& \mathbf{T}_v^T\mathbf{T}_1 - u^{-1}T_{v\cdot}T_{1\cdot} & \cdots & \mathbf{T}_v^T\mathbf{T}_v - u^{-1}T_{v\cdot}^2
\end{vmatrix}\label{supp::le2.1}\\
&= (1-c)u \cdot |\mathbf{T}^T\mathbf{T} - u^{-1}\mathbf{T}^T\mathbf{1}_u\mathbf{1}_u^T\mathbf{T}|\notag\\
&\leqslant (1-c)u \cdot |\mathbf{T}^T\mathbf{T}|\label{supp::le2.2}\\
&\leqslant (1-c)u^{v+1}\label{supp::le2.3},
\end{align}
where the equality (\ref{supp::le2.1}) is obtained by the elementary transformation of the determinant, the equality in (\ref{supp::le2.2}) holds if and only if the sum of the columns of $\mathbf{T}$ is zero, and the equality in (\ref{supp::le2.3}) holds if and only if $\mathbf{T}$ forms a two-level OA with $u$ runs by Lemma \ref{supp::lem1}.

Therefore, $|\tilde{\mathbf{T}}^{T}\tilde{\mathbf{T}} - cu^{-1}\tilde{\mathbf{T}}^{T}\mathbf{1}_u\mathbf{1}_u^{T}\tilde{\mathbf{T}}| = (1-c)u^{v+1}$ if and only if $\mathbf{T}$ forms a two-level OA with $u$ runs. This completes the proof.

\vspace{10pt}
\textbf{Proof of Lemma \ref{lem0}} Note that
\begin{align*}
|\mathbf{M}^*_{i}|
&= |\mathbf{X}_i^{*T}\mathbf{V}_i^{*-1}\mathbf{X}_i^*| \\
&= \sigma_E^{-2p} \cdot |\mathbf{X}_i^{*T}\mathbf{X}_i^* - (1-\gamma_i) n_i^{-1}\mathbf{X}_i^{*T}\mathbf{1}_{n_i}\mathbf{1}_{n_i}^{T}\mathbf{X}_i^*|,
\end{align*}
where the last equality in the above decomposition comes from the Woodbury formula \citep{horn2012matrix}, that is $\mathbf{V}_i^{*-1} = (\mathbf{I}_{n_{i} } - (1-\gamma_i)n_{i} ^{-1}\mathbf{1}_{n_{i} }\mathbf{1}_{n_{i} }^T)/\sigma^2_E$, where $\gamma_i = \sigma^2_E/(\sigma^2_E + n_{i} \sigma^2_A)$.

Then the desired result follows directly from Lemma \ref{supp::lem2}.

\vspace{10pt}
\textbf{Proof of Throrem \ref{th1.2}} From Hadamard inequality,
\begin{align}
|\mathbf{M}^*|
&= \frac{1}{\sigma_E^{2p}} \begin{vmatrix}
  \sum_{i=1}^{R} \gamma_in_i &  \sum_{i=1}^{R} \gamma_i\mathbf{1}_{n_i}^T\mathbf{Z}^*_i  \\
  \sum_{i=1}^{R} \gamma_i\mathbf{Z}_i^{*T}\mathbf{1}_{n_i} & \sum_{i=1}^{R} \mathbf{Z}_i^{*T}\mathbf{V}_i^{*-1}\mathbf{Z}^*_i
\end{vmatrix}\notag\\
& \leqslant \frac{1}{\sigma_E^{2p}} \left[\sum_{i=1}^{R} \gamma_in_i\right] \cdot \prod_{k=1}^{p-1} \left[\sum_{i=1}^{R} \left(\mathbf{Z}_{i.k}^{*T}\mathbf{Z}_{i.k}^{*}- \frac{(1-\gamma_i)}{n_i}\mathbf{Z}_{i.k}^{*T}\mathbf{1}_{n_i}\mathbf{1}_{n_i}^T\mathbf{Z}_{i.k}^{*}\right)\right], \label{supp::pf1.2}
\end{align}
where $\mathbf{Z}^*_{i.k}$ is the $k$th column of $\mathbf{Z}^*_i$, and the equality in (\ref{supp::pf1.2}) holds if and only if $\mathbf{M}^*$ is a diagonal matrix.

Note that $\mathbf{M}^*$ becomes a diagonal matrix when the subdata design matrix $\mathbf{Z}^*_i$ of each group forms an OA. At this time, $|\mathbf{M}^*|$ can reaches the maximum $\sigma_E^{-2p}\left[\sum_{i=1}^{R}\gamma_in_i\right] n^{p-1}$. This completes the proof of the first result.

Note that $f(x) = x/(\sigma_E^2 + x \sigma_A^2)$ is the concave function on $[1,n]$. For any $n_1, n_2,\ldots,n_R \in [1,n]$, we have
\begin{align}
\sum_{i=1}^{R}\frac{n_i}{\sigma_E^2 + n_i\sigma_A^2} = \sum_{i=1}^{R} f(n_i) \leqslant R f\left(\frac{\sum_{i=1}^{R}n_i}{R}\right) = \frac{nR}{R\sigma_E^2 + n\sigma_A^2}, \label{supp::pf1.3}
\end{align}
by Jensen inequality, and the equality in (\ref{supp::pf1.3}) holds if and only if $n_1 = \ldots = n_R$. The desired result holds directly.

\vspace{10pt}
\textbf{Proof of Theorem \ref{co1}}
Let $\mathbf{M}_{11}^* = \sum_{i=1}^{R} \gamma_in_i$, $\mathbf{M}_{12}^* = \sum_{i=1}^{R} \gamma_i\mathbf{1}_{n_i}^T\mathbf{Z}^*_i$, $\mathbf{M}_{21}^* = \sum_{i=1}^{R} \gamma_i\mathbf{Z}_i^{*T}\mathbf{1}_{n_i}$, $\mathbf{M}_{22}^* = \sum_{i=1}^{R} \mathbf{Z}_i^{*T}\mathbf{V}_i^{*-1}\mathbf{Z}^*_i$ and
\begin{align*}
 \mathbf{M}_{22.1}^* &= \mathbf{M}_{22}^* - \mathbf{M}_{21}^* \mathbf{M}_{11}^{*-1} \mathbf{M}_{12}^*  \\
 &= \sum_{i=1}^{R}\mathbf{Z}_i^{*T}\mathbf{V}_i^{*-1}\mathbf{Z}_i^* - \left(\sum_{i=1}^{R} \gamma_i\mathbf{Z}_i^{*T}\mathbf{1}_{n_i}\right)\left(\frac{1}{\sum_{i=1}^{R} \gamma_in_i}\right)\left(\sum_{i=1}^{R} \gamma_i\mathbf{1}_{n_i}^T\mathbf{Z}^*_i\right).
\end{align*}
By inverting $\mathbf{M}^*$
we can obtain that
\begin{align*}
\mathbf{M}^{*-1} = \begin{pmatrix}
 \mathbf{M}_{11}^{*-1} + \mathbf{M}_{11}^{*-1}\mathbf{M}_{12}^{*}\mathbf{M}_{22.1}^{*-1}\mathbf{M}_{21}^{*} \mathbf{M}_{11}^{*-1}& -\mathbf{M}_{11}^{*-1}\mathbf{M}_{12}^{*}\mathbf{M}_{22.1}^{*-1}\\
 -\mathbf{M}_{22.1}^{*-1}\mathbf{M}_{21}^{*} \mathbf{M}_{11}^{*-1} & \mathbf{M}_{22.1}^{*-1}
\end{pmatrix}.
\end{align*}

For $\mathbf{M}_{22.1}^{*-1}$, we have
\begin{align}
\mathrm{tr}(\mathbf{M}_{22.1}^{*-1}) \geqslant \mathrm{tr}(\mathbf{M}_{22}^{*-1}). \label{supp::pf.co1.new}
\end{align}
When the design matrix of each group $\mathbf{Z}^*_i$ forms an OA, the equality in \eqref{supp::pf.co1.new} holds, i.e.,
\begin{align}
\mathrm{tr}(\mathbf{M}_{22.1}^{*-1}) =  \frac{(p-1)\sigma_E^2}{n}. \label{supp::pf.co1.1}
\end{align}
We have
\begin{align}
&\mathrm{tr}(\mathbf{M}^{*-1}) \notag \\
=& \mathrm{tr}(\mathbf{M}_{11}^{*-1}) + \mathrm{tr}(\mathbf{M}_{11}^{*-1}\mathbf{M}_{12}^{*}\mathbf{M}_{22.1}^{*-1}\mathbf{M}_{21}^{*} \mathbf{M}_{11}^{*-1}) + \mathrm{tr}(\mathbf{M}_{22.1}^{*-1})
\notag\\
= & \sigma_E^{2}\left[
\frac{1}{\sum_{i=1}^{R} \gamma_in_i} + \left(\frac{1}{\sum_{i=1}^{R} \gamma_in_i}\right)\left(\sum_{i=1}^{R} \gamma_i\mathbf{1}_{n_i}^T\mathbf{Z}^*_i\right)\mathbf{M}_{22.1}^{*-1}\left(\sum_{i=1}^{R} \gamma_i\mathbf{Z}_i^{*T}\mathbf{1}_{n_i}\right)\left(\frac{1}{\sum_{i=1}^{R} \gamma_in_i}\right)+ \frac{p-1}{n}\right] \notag\\
=& \sigma_E^2\left(\frac{1}{\sum_{i=1}^{R}\gamma_in_i} + \frac{p-1}{n}\right),\notag
\end{align}
when the design matrix of each group $\mathbf{Z}^*_i$ forms an OA,
and the equality in (\ref{co1.1}) is proved.

The equation of (\ref{co1.2}) has been proved in (\ref{supp::pf1.3}). This completes the proof.

\vspace{10pt}
Let $\mathbf{A} = \mathrm{diag}((n\gamma)^{-1}, n^{-1}, \ldots, n^{-1})$ is a $p \times p$ diagonal matrix with $n = Rm$ and $\gamma = \sigma^2_E/(\sigma^2_E + m \sigma^2_A)$.
For $i = 1,2,\ldots,R$, let $\tilde{\mathbf{L}}_i^* = (\mathbf{1}, {\mathbf{L}}_i^*) = (\mathbf{1}, \mathbf{L}_{i2}^*,\ldots,\mathbf{L}_{ip}^*)$ and $\tilde{\mathbf{D}}_i^* = (\mathbf{0}, {\mathbf{D}}_i^*) = (\mathbf{1}, \mathbf{D}_{i2}^*,\ldots,\mathbf{D}_{ip}^*)$.
The following two lemmas are needed in the proof of Theorem \ref{th4}.

\begin{lemma}\label{supp::lem3}
 Let $\tilde{\mathbf{L}}^* = (\tilde{\mathbf{L}}_{1}^{*T}, \ldots, \tilde{\mathbf{L}}_{R}^{*T})^T$ and $\tilde{\mathbf{D}}^* = (\tilde{\mathbf{D}}_{1}^{*T}, \ldots,\tilde{\mathbf{D}}_{R}^{*T})^T$.
 Assume that $||\tilde{\mathbf{D}}^*||_{\infty} = o(1)$ as $m \rightarrow \infty$, for $i = 1,2,\ldots,R$. Then as $m \rightarrow \infty$,
 \begin{enumerate}
                  \item[(1)] $||\mathbf{A}\tilde{\mathbf{L}}^{*T}\mathbf{V}^{*-1}\tilde{\mathbf{D}}^*||_\infty=o(1)$, $||\mathbf{A}\tilde{\mathbf{D}}^{*T}\mathbf{V}^{*-1}\tilde{\mathbf{L}}^*||_\infty = o(1)$ and $||\mathbf{A}\tilde{\mathbf{D}}^{*T}\mathbf{V}^{*-1}\tilde{\mathbf{D}}^*||_\infty = o(1)$.
                  \item[(2)] $\left[\mathbf{A}(\mathbf{X}^{*T}\mathbf{V}^{*-1}\mathbf{X}^{*})\right]^{-1} = \sigma_E^2\mathbf{I}_p + o(1)$.
       \end{enumerate}
\end{lemma}

\textbf{Proof}$(1).$
From the orthogonality of OA, we have
\begin{align*}
\tilde{\mathbf{D}}_{i}^{*T}\mathbf{V}_{i}^{*-1}\tilde{\mathbf{L}}_i^* &=
\frac{1}{\sigma_E^2}\begin{pmatrix}
 \mathbf{0}^T_{m\times 1}\\
 \mathbf{D}_{i2}^{*T}\left[\mathbf{I}_m - \frac{(1-\gamma)}{m}\mathbf{1}_m\mathbf{1}_m^T\right]\\
 \vdots\\
 \mathbf{D}_{ip}^{*T}\left[\mathbf{I}_m - \frac{(1-\gamma)}{m}\mathbf{1}_m\mathbf{1}_m^T\right]\\
\end{pmatrix}
\times
\begin{pmatrix}
 &\mathbf{1} & \mathbf{L}_{i2}^* & \cdots  &\mathbf{L}_{ip}^*
\end{pmatrix}\\
&=
\frac{1}{\sigma_E^2}
\begin{pmatrix}
 0 &  0 & \cdots & 0\\
  \gamma\mathbf{D}_{i2}^{*T}\mathbf{1}_m & \mathbf{D}_{i2}^{*T}\mathbf{L}_{i2}^* & \cdots & \mathbf{D}_{i2}^{*T}\mathbf{L}_{ip}^*\\
 \vdots & \vdots & & \vdots\\
 \gamma\mathbf{D}_{ip}^{*T}\mathbf{1}_m & \mathbf{D}_{ip}^{*T}\mathbf{L}_{i2}^* & \cdots &\mathbf{D}_{ip}^{*T}\mathbf{L}_{ip}^*
\end{pmatrix}.
\end{align*}
where $\gamma = \sigma_E^2/(\sigma_E^2 + m\sigma_A^2)$, which converges to $0$ as $m \rightarrow \infty$.

Then, from the assumption $||\tilde{\mathbf{D}}^*||_{\infty} = o(1)$ and the definition of $\mathbf{A}$, $||\tilde{\mathbf{D}}_i^*||_{\infty} =  o(1)$, and the elements of $\mathbf{A}\tilde{\mathbf{D}}^{*T}\mathbf{V}^{*-1}\tilde{\mathbf{L}}^* = \mathbf{A}\sum_{i=1}^{R}\tilde{\mathbf{D}}_i^{*T}\mathbf{V}_i^{*-1}\tilde{\mathbf{L}}_i^*$ converge
to $0$ as $m \rightarrow \infty$.

Similar arguments can prove that
$\mathbf{A}\tilde{\mathbf{L}}^{*T}\mathbf{V}^{*-1}\tilde{\mathbf{D}}^* = o(1)$, $\mathbf{A}\tilde{\mathbf{D}}^{*T}\mathbf{V}^{*-1}\tilde{\mathbf{D}}^* = o(1)$,
as $m \rightarrow \infty$.

$(2).$ Note that
\begin{align*}
[\mathbf{A}(\mathbf{X}^{*T}\mathbf{V}^{*-1}\mathbf{X}^{*})]^{-1} = [\mathbf{A}(\tilde{\mathbf{L}}^{*T}\mathbf{V}^{*-1}\tilde{\mathbf{L}}^* + \tilde{\mathbf{L}}^{*T}\mathbf{V}^{*-1}\tilde{\mathbf{D}}^* + \tilde{\mathbf{D}}^{*T}\mathbf{V}^{*-1}\tilde{\mathbf{L}}^* + \tilde{\mathbf{D}}^{*T}\mathbf{V}^{*-1}\tilde{\mathbf{D}}^*)]^{-1}.
\end{align*}

From the orthogonality between any two columns of OA, we have
\begin{align*}
[\mathbf{A}(\mathbf{X}^{*T}\mathbf{V}^{*-1}\mathbf{X}^{*})]^{-1} = [\mathbf{A}(\tilde{\mathbf{L}}^{*T}\mathbf{V}^{*-1}\tilde{\mathbf{L}}^* + o(1))]^{-1} = \sigma_E^2\mathbf{I}_p + o(1), m \rightarrow \infty.
\end{align*}

\begin{lemma}\label{supp::lem4}(Theorem 2.7.3, \cite{lehmann2004elements})
Let random variables $\xi_{1}, \xi_{2}, \ldots, \xi_{n}$ be i.i.d with $\mathrm{E}(\xi_i) = 0$, $Var(\xi_i) = \sigma^2 > 0$, and $\mathrm{E}|\xi_i^3| < \infty$. $g_{1}, g_{2}, \ldots,g_{n}$ are real numbers and not all zero. Then
\begin{align*}
\frac{\sum_{i=1}^{n}g_{i}\xi_i}{\sigma\sqrt{\sum_{i=1}^{n}g_{i}^2}} \stackrel{d}{\longrightarrow} N(0, 1),
\end{align*}
provided
\begin{align}
\underset{i=1,\ldots,n}{\max} \left(g_{i}^2\right) = o\left(\sum_{i=1}^{n}g_{i}^2\right). \label{supp::lem4.1}
\end{align}
\end{lemma}

\vspace{10pt}
\textbf{Proof of Theorem \ref{th4}}
 We fist prove $\hat{\boldsymbol{\beta}}^* = (\mathbf{X}^{*T}\mathbf{V}^{*-1}\mathbf{X}^*)^{-1}\mathbf{X}^{*T}\mathbf{V}^{*-1}\mathbf{Y}^*$ satisfies Theorem \ref{th4}.

By the definition of $\hat{\boldsymbol{\beta}}^*$, it yields that $\mathrm{E}(\hat{\boldsymbol{\beta}}^*) = \boldsymbol{\beta}$.

Let $\mathbf{\eta}^* = \mathbf{a}^* + \mathbf{e}^*$. Note that
\begin{align*}
  \hat{\boldsymbol{\beta}}^*
  &= (\mathbf{X}^{*T}\mathbf{V}^{*-1}\mathbf{X}^{*})^{-1}\mathbf{X}^{*T}\mathbf{V}^{*-1}(\mathbf{X}^{*}\boldsymbol{\beta} + \mathbf{\eta}^*)\\
  &= \boldsymbol{\beta} + (\mathbf{X}^{*T}\mathbf{V}^{*-1}\mathbf{X}^{*})^{-1}\mathbf{A}^{-1}\mathbf{A}\mathbf{X}^{*T}\mathbf{V}^{*-1}\mathbf{\eta}^*\\
  &= \boldsymbol{\beta} + \left[\mathbf{A}(\mathbf{X}^{*T}\mathbf{V}^{*-1}\mathbf{X}^{*})\right]^{-1}\mathbf{A}\mathbf{X}^{*T}\mathbf{V}^{*-1}\mathbf{\eta}^*.
\end{align*}
Thus,
$$\hat{\boldsymbol{\beta}}^* - \boldsymbol{\beta} = \left[\mathbf{A}(\mathbf{X}^{*T}\mathbf{V}^{*-1}\mathbf{X}^{*})\right]^{-1}\mathbf{A}\mathbf{X}^{*T}\mathbf{V}^{*-1}\mathbf{\eta}^*.$$

We have proved that $\left[\mathbf{A}(\mathbf{X}^{*T}\mathbf{V}^{*-1}\mathbf{X}^{*})\right]^{-1} = \sigma_E^2\mathbf{I}_p + o(1)$, as $m \rightarrow \infty$ in Lemma \ref{supp::lem3}. Next, we will prove that $\mathbf{A}\mathbf{X}^{*T}\mathbf{V}^{*-1}\mathbf{\eta}^* = \mathbf{A}(\tilde{\mathbf{L}}^{*T}\mathbf{V}^{*-1}\mathbf{\eta}^* +\tilde{\mathbf{D}}^{*T}\mathbf{V}^{*-1}\mathbf{\eta}^*)$ is asymptotically normal.
By the fact that $\mathbf{\eta}^* = \mathbf{a}^* + \mathbf{e}^* = O_p(1)$ and $||\tilde{\mathbf{D}}_i^*||_{\infty} = o(1)$, we have
\begin{align*}
\mathbf{A}\tilde{\mathbf{D}}^{*T}\mathbf{V}^{*-1}\mathbf{\eta}^* = \mathbf{A}\sum_{i=1}^{R}\tilde{\mathbf{D}}_i^{*T}\mathbf{V}_{i}^{*-1}\mathbf{\eta}_i^* = o_p(1),
\end{align*}
as $n \rightarrow \infty$ and a fixed $R$. Therefore, we only need to prove that $\mathbf{A}\tilde{\mathbf{L}}^{*T}\mathbf{V}^{*-1}\mathbf{\eta}^*$ is asymptotically normal.

Let $\hat{\boldsymbol{\beta}}^{*} = (\hat{\beta}_{1}^{*},\hat{\boldsymbol{\beta}}_{-1}^{*T})^T$. We next prove that the joint distribution of the remaining $p-1$ elements of $\mathbf{A}\tilde{\mathbf{L}}^{*T}\mathbf{V}^{*-1}\mathbf{\eta}^*$ follows a multivariate normal distribution.
From Cramer-wold devive, it is only necessary to prove that any linear combination of these elements follows an univariate normal distribution, i.e., for all constant $\mathbf{c} = (c_2, \ldots,c_p)^T$, $\sqrt{n}\sum_{k=2}^{p}c_k(\mathbf{A}\tilde{\mathbf{L}}^{*T}\mathbf{V}^{*-1}\mathbf{\eta}^*)_k$ follows a univariate normal distribution.

Let $L_{ijk}$ be the $(j, k)$-th entry of $\tilde{\mathbf{L}}^{*}$ and observing that
\begin{align*}
\sqrt{n}\sum_{k=2}^{p}c_k(\mathbf{A}\tilde{\mathbf{L}}^{*T}\mathbf{V}^{*-1}\mathbf{\eta}^*)_k
&= \frac{1}{\sqrt{n}\sigma_E^{2}} \sum_{k=2}^{p}c_k \sum_{i=1}^{R}\sum_{j=1}^{m}L_{ijk}e_{ij}\\
&= \frac{1}{\sqrt{n}\sigma_E^{2}} \sum_{i=1}^{R}\sum_{j=1}^{m}\left(\sum_{k=2}^{p}c_k L_{ijk}\right)e_{ij}.
\end{align*}
Let $g_{ij} = \sum_{k=2}^{p}c_k L_{ijk}$, and based on Lemma \ref{supp::lem4}, we only need to verify
\begin{align*}
\underset{{i=1,\ldots,R\atop j=1,\ldots,m}}{\max} \left(g_{ij}^2\right) = o\left(\sum_{i=1}^{R}\sum_{j=1}^{m}g_{ij}^2\right).
\end{align*}
Note that $L_{ijk}=\pm1$ and the orthogonality of $\mathbf{L}^{*}$,  we have
\begin{align*}
&g_{ij}^2 = \left(\sum_{k=2}^{p}c_k L_{ijk}\right)^2 \leqslant (p-1) ||\mathbf{c}||^2,\\
& \sum_{i=1}^{R}\sum_{j=1}^{m} g_{ij}^2  =\mathbf{c}^T\mathbf{L}^{*T}\mathbf{L}^*\mathbf{c}=n  ||\mathbf{c}||^2,
\end{align*}
 and then  $\hat{\boldsymbol{\beta}}^* = (\mathbf{X}^{*T}\mathbf{V}^{*-1}\mathbf{X}^*)^{-1}\mathbf{X}^{*T}\mathbf{V}^{*-1}\mathbf{Y}^*$ satisfies Theorem  \ref{th4}.

The moment estimators $\hat{\sigma}_A^2$ and $\hat{\sigma}_E^2$ are consistent estimators (Theorem 5, \cite{gao2019estimation}), and then the conclusion in Theorem \ref{th4} is also valid for $$\check{\boldsymbol{\beta}}^* = (\mathbf{X}^{*T}\mathbf{\hat{V}}^{*-1}\mathbf{X}^*)^{-1}\mathbf{X}^{*T}\mathbf{\hat{V}}^{*-1}\mathbf{Y}^*$$  based on Slutsky theorem. We complete the poof.

\section{Additional numerical results}\label{appB}

\subsection{Estimation of intercept} \label{appB.1}

\begin{figure}[htbp]
		\centering
\includegraphics[width=18cm]{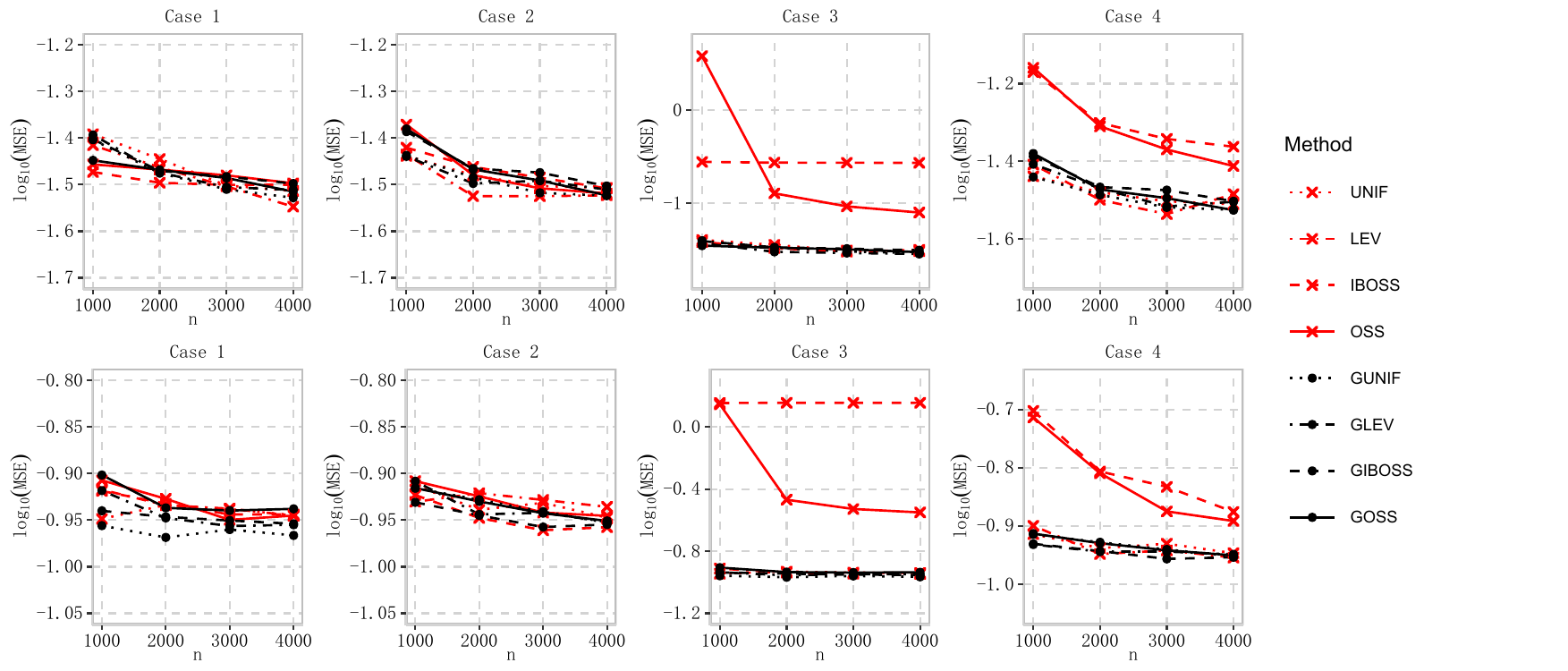}
	\caption{$\mathrm{log}_{10}$(MSE) of $\breve{\beta}_{1}^{*}$ for different subdata size $n$. The upper panels are for $a_{i}\sim N(0, 0.5)$, and the lower panels for $a_{i}\sim t(3)$. The full data size is $N=1.5\times 10^5$.}
	\label{supp::figS1}
\end{figure}

To demonstrate the efficiency of the proposed GOSS for estimating the intercept term, we calculate the mean square error of the intercept parameter $\mathrm{MSE}_{\beta_1}$ under different subsampling methods. That is
\begin{equation*}
\mathrm{MSE}_{\beta_1} = B^{-1}\sum_{b =1}^{B} (\breve{\beta}_{1}^{*(b)} - \beta_{1})^2,
\end{equation*}
where $\breve{\beta}_{1}^{*(b)}$ is the generalized least squares (GLS) estimator of $\beta_{1}$ based on subdata in the $b$th repetition.
From Figure \ref{supp::figS1} below we can see that overall all the subsampling methods have similar performances in estimating the intercept parameter. They typically have small estimation errors, except for Cases 3 and 4 where IBOSS and OSS perform worse than other methods and have relatively bigger estimation errors. Note that both IBOSS and OSS use a modified intercept estimator for linear regression, see \cite{wang2019information} and \cite{WangL}, but such a modification is unclear for linear mixed models.


\subsection{Prediction of full data response} \label{appB.2}

\begin{figure}[htbp]
		\centering
\includegraphics[width=18cm]{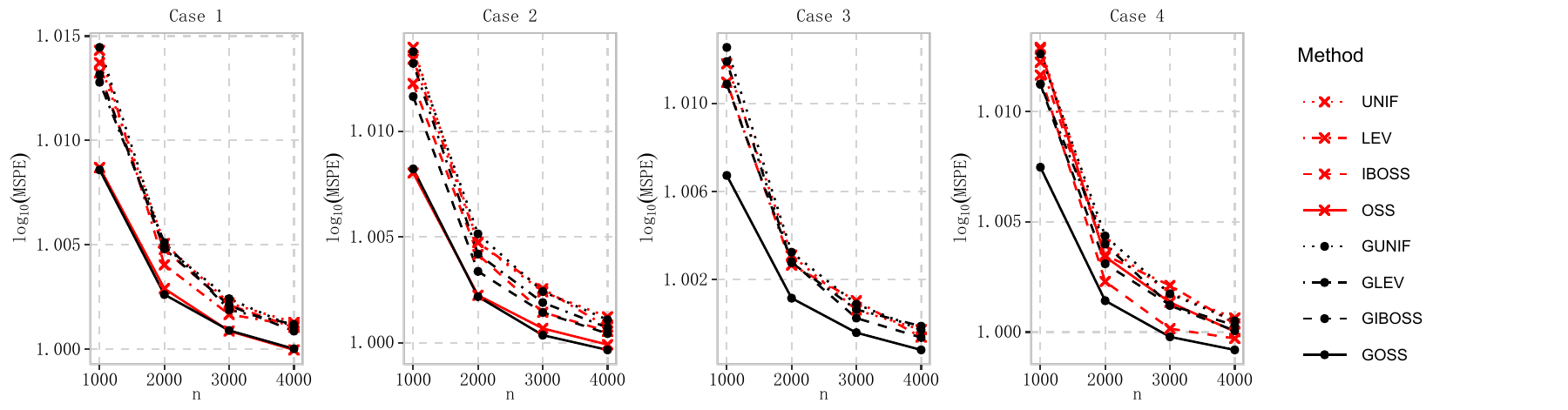}
	\caption{$\mathrm{log}_{10}$(MSPE) of $\hat{y}_{ij}$ for different subdata size $n$ when $a_{i}\sim N(0, 0.5)$. The  panels are for all eight subsampling methods, and the third panel for the seven methods except OSS. The full data size is $N=1.5\times 10^5$.}
	\label{supp::figS2}
\end{figure}

We evaluate the prediction performance of different subsampling methods by comparing the mean square prediction error $(\mathrm{MSPE})$ over the full data, that is,
$$\mathrm{MSPE} = \frac{1}{N} \sum_{i=1}^{R} \sum_{j=1}^{C_i} (y_{ij} - \hat{y}_{ij} )^2,$$ where $\hat{y}_{ij} = \mathbf{x}_{ij}^{T}\hat{\boldsymbol{\beta}}^* + \hat{a}_{i}^*$, $\hat{\boldsymbol{\beta}}^*$ is GLS estimator based on the subdata as defined in (2), $\hat{a}_{i}^*$ is the prediction of $a_{i}$ based on subdata \citep{henderson1950estimation, henderson1963selection}, which is calculated by
$$\hat{a}_{i}^* = \frac{\hat{\sigma}_A^2}{\hat{\sigma}_E^2 + n_i\hat{\sigma}_A^2} \sum_{j=1}^{n_i} (y_{ij}^* - \mathbf{x}_{ij}^{*T}\hat{\boldsymbol{\beta}}^*),$$ where $\hat{\sigma}_A^2$ and $\hat{\sigma}_E^2$ are the consistent estimators of $\sigma_A^2$ and $\sigma_E^2$ based on the subdata.
Figure \ref{supp::figS2} plots the prediction performance of the different methods when $a_{i}\sim N(0, 0.5)$.
Given the significantly difference in MSPE between OSS and other methods in Case 3, we only presented the comparison of the other seven methods.
We can find that the prediction performance of GOSS overall outperforms the other methods, especially in Cases 3 and 4 where the data across different groups come from different distributions.

\subsection{Estimation in the presence of model misspecification} \label{appB.3}

To show the robustness of the GOSS estimator under various misspecification terms, we add the model misspecification term in the model and evaluate the performance of the estimation of the slope parameter by MSE in \eqref{mse1}. Specifically, we assume the data from the model
\begin{equation*}
    y_{ij} = \mathbf{x}_{ij}^{T}\boldsymbol{\beta} + h(\mathbf{x}_{ij}) + a_i + e_{ij},
\end{equation*}
where the settings of $\boldsymbol{\beta}$ and $e$ are the same as in the Simulation studies, $a_{i}\sim N(0, 0.5)$ and
$h$ is the model misspecification.
For $\mathbf{x}_{ij}$ generated by Case 1 - Case 4, the $h$ we consider here are special cases of the misspecification terms considered in \cite{meng2021lowcon}, and they are
\begin{description}
    \item[H1.] $h(\mathbf{x}_{ij}) = 0.1 x_{ij1}x_{ij2}$;
    \item[H2.] $h(\mathbf{x}_{ij}) = 0.1 x_{ij1}\sin(x_{ij2})$.
\end{description}

\begin{figure}[htbp]
		\centering
\includegraphics[width=18cm]{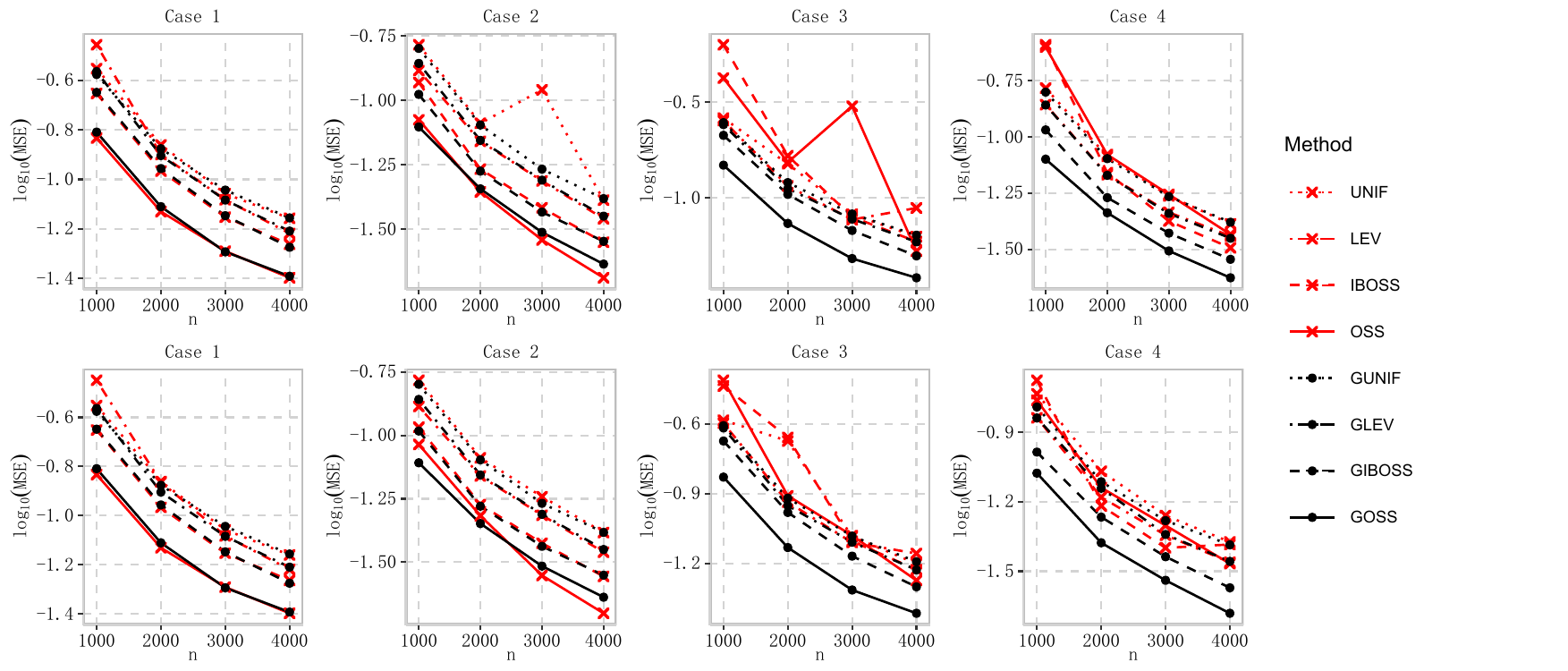}
	\caption{$\mathrm{log}_{10}$(MSE) of the estimated slope parameters for different subdata size $n$ when there is a  misspecification term in the model. The upper panels are for H1, and the lower panels for H2. $a_{i}\sim N(0, 0.5)$ and the full data size is $N=1.5\times 10^5$.}
	\label{supp::figS3}
\end{figure}

In the following, we give the performance of the subsampling method with respect to slope parameter estimation under the misspecification terms H1, and H2. From Figure \ref{supp::figS3}, We can find that for all the misspecifications considered, GOSS has the best estimation performance in almost all cases, especially for Cases 3 and 4 where the data come from different distributions.

\subsection{Estimation results of $\hat{\sigma}_A^2$ and $\hat{\sigma}_E^2$}
\label{appB.4}

Below we give the estimation performance of $\hat{\sigma}_A^2$ and $\hat{\sigma}_E^2$ for the simulation settings of Section \ref{sec4} using the above estimation methods.

\begin{figure}[htbp]
\centering
\includegraphics[width=18cm]{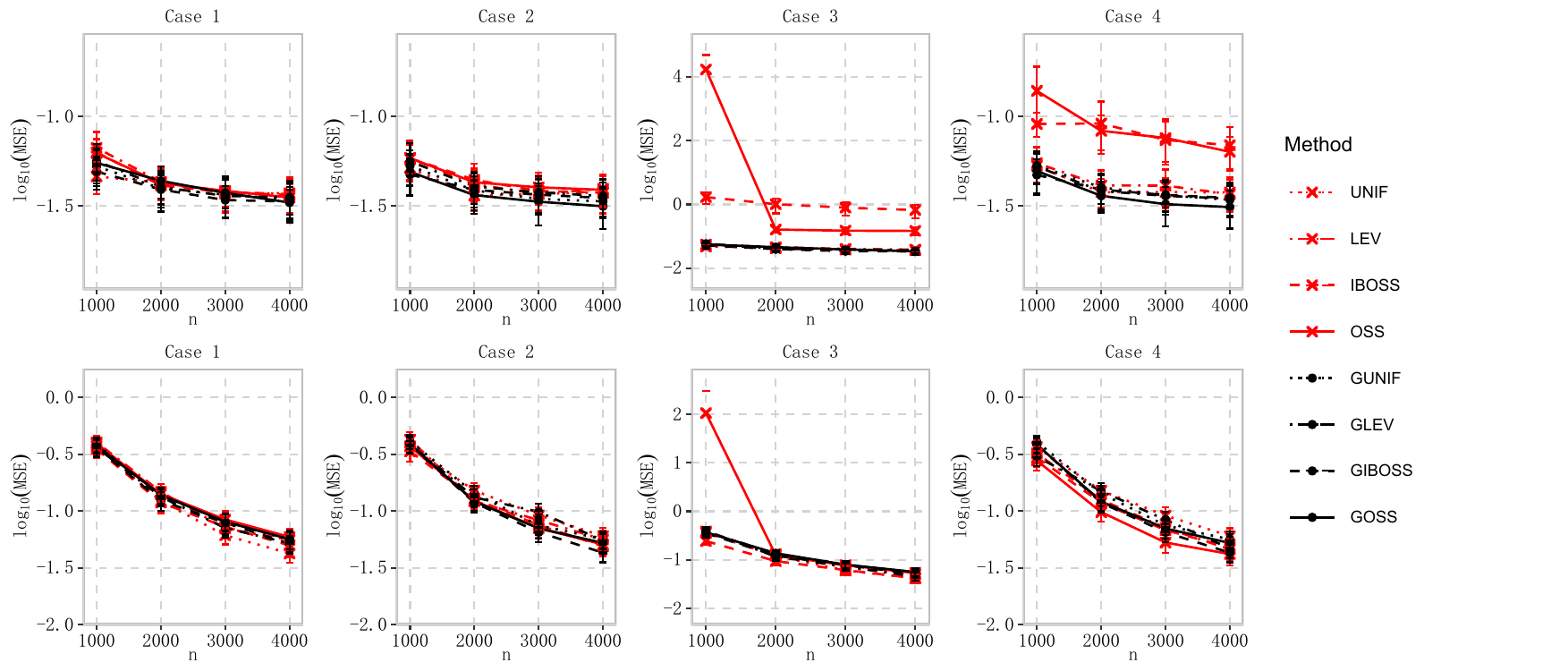}
	\caption{$\mathrm{log}_{10}$(MSE) of $\hat{\sigma}^2_A$ and $\hat{\sigma}^2_E$ for different subdata size $n$ when $a_{i}\sim N(0, 0.5)$. The upper panels are for $\hat{\sigma}^2_A$, and the lower panels for $\hat{\sigma}^2_E$. The full data size is $N=1.5\times 10^5$. The bars represent standard errors obtained from $200$ replicates.}
	\label{supp::fig1}
\end{figure}

\begin{figure}[htbp]
\centering
\includegraphics[width=18cm]{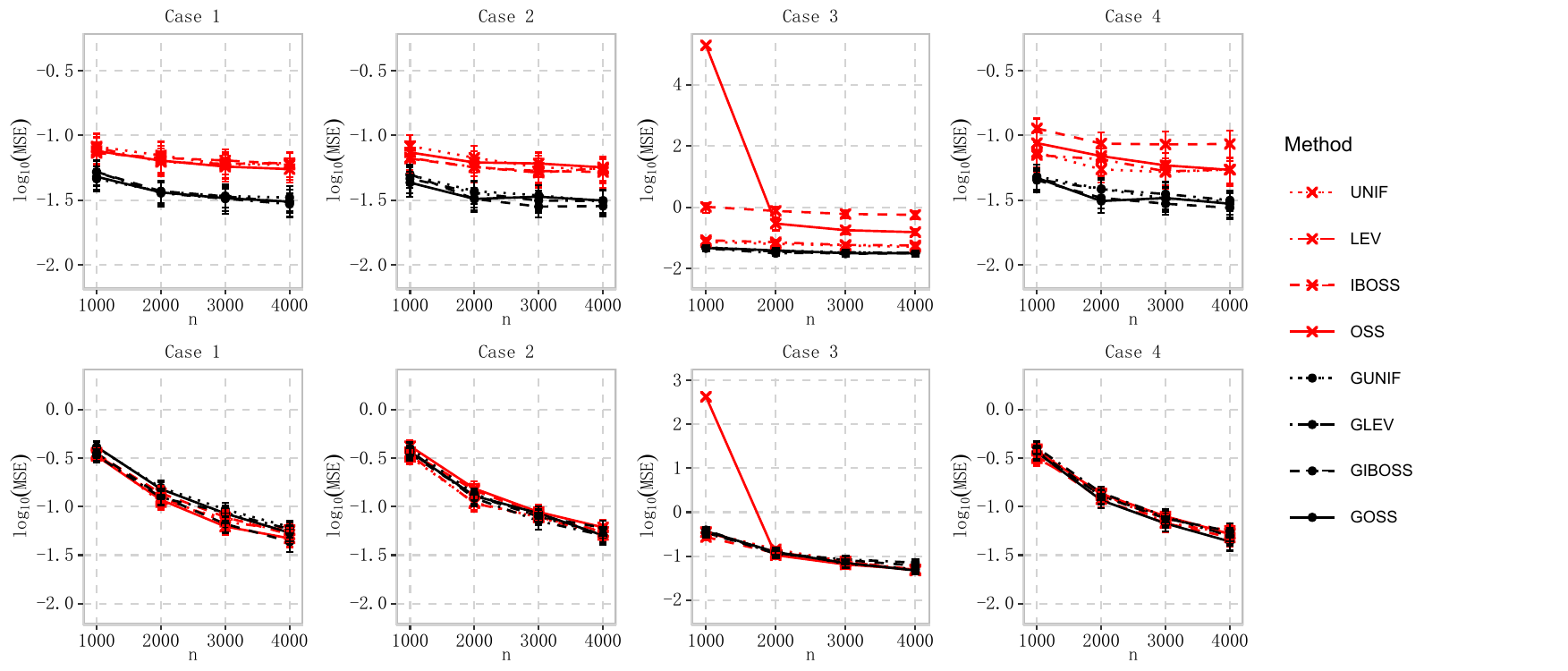}
	\caption{$\mathrm{log}_{10}$(MSE) of $\hat{\sigma}^2_A$ and $\hat{\sigma}^2_E$ for different subdata size $n$ when $a_{i}\sim N(0, 0.5)$. The upper panels are for $\hat{\sigma}^2_A$, and the lower panels for $\hat{\sigma}^2_E$. The full data size is $N=5.5\times 10^5$. The bars represent standard errors obtained from $200$ replicates.}
	\label{supp::figC}
\end{figure}

\section{The OSS algorithm}\label{appC}


For the $i$th group, we use the OSS for subsampling.
Specifically, OSS searches for the subdata $\left\{\mathbf{Z}_{i}^*, \mathbf{Y}_{i}^*\right\}$ that minimize the discrepancy function:
\begin{align*}
L\left(\mathbf{Z}_{i}^*\right)=\sum_{1 \leqslant j<j' \leqslant n_i}\left[(p-1)-\left\|\mathbf{z}_{ij}^*\right\|^{2} / 2-\left\|\mathbf{z}_{ij'}^*\right\|^{2} / 2+\delta\left(\mathbf{z}_{ij}^*, \mathbf{z}_{ij'}^*\right)\right]^{2},
\end{align*}
where $$\delta\left(\mathbf{z}_{ij}^*, \mathbf{z}_{ij'}^*\right)=\sum_{k=2}^{p} \delta_{1}\left(x_{ijk}^*, x_{ij'k}^{*}\right),$$
and $\delta_{1}(x, y)$ is $1$ if both $x$ and $y$ have the same sign and $0$ otherwise.
Assume the algorithm is at the $j$th iteration where $\mathbf{Z}_{ij}^*$ is the new matrix obtained by adding $\mathbf{z}_{ij}^*$ to $\mathbf{Z}_{i(j-1)}^*$, $j = 2, \ldots, m$. To select next point $\mathbf{z}_{ij}^*$, OSS aims to minimise the discrepancy:
\begin{align}\label{supp::algnew1}
l\left(\mathbf{z}_{ij}|\mathbf{Z}_{i(j-1)}^*\right)=\sum_{\mathbf{z}_{ij'}^*\in \mathbf{Z}_{i(j-1)}^*}\left[(p-1)-\left\|\mathbf{z}_{ij}\right\|^{2} / 2-\left\|\mathbf{z}_{ij'}^*\right\|^{2} / 2+\delta\left(\mathbf{z}_{ij}^*, \mathbf{z}_{ij'}^*\right)\right]^{2},
\end{align}
which is the discrepancy introduced by adding  $\mathbf{z}_{ij}^*$ to $\mathbf{Z}_{i(j-1)}^*$.

A key advantage of the discrepancy function in (\ref{supp::algnew1}) is that it allows sequential minimization to speed up the search. After $\mathbf{z}_{i(j-1)}^*$ is selected, it only need to computer  $l\left(\mathbf{z}_{ij} \mid \mathbf{z}_{i(j-1)}^* \right)$ to select the next data point $\mathbf{z}_{ij}^*$, where
   \begin{equation}\label{supp::algnew2}
   l\left(\mathbf{z}_{ij} \mid \mathbf{z}_{i(j-1)}^* \right)=\left[(p-1)-\|\mathbf{z}_{ij}\|^{2} / 2-\left\|\mathbf{z}_{i(j-1)}^*\right\|^{2} / 2+\delta\left(\mathbf{z}_{ij}, \mathbf{z}_{i(j-1)}^*\right)\right]^{2},
     \end{equation}
and the computational complexity of \eqref{supp::algnew2} is $O(Np)$.
To further reduce the computation, OSS deletes some data points in  $\mathbf{Z}_{i}$  with large values of  $l\left(\mathbf{z}_{ij}|\mathbf{Z}_{i(j-1)}^*\right)$  so that these points will not be considered at the  $(j+1)$th iteration.
Algorithm \ref{supp::algS1} outlines the steps of using OSS in the $i$th group.

\begin{algorithm}[htpt]
\caption{OSS algorithm for group $i$}
\begin{algorithmic}\label{supp::algS1}
\REQUIRE
Full data $\left\{\mathbf{Z}_i, \mathbf{Y}_i\right\}$,  subdata size $m = n/R$
\ENSURE  The subdata $\left\{\mathbf{Z}_{i}^*, \mathbf{Y}_{i}^*\right\}$
\STATE Set $\left\{\mathbf{Z}_{i1}^*, \mathbf{Y}_{i1}^*\right\} \leftarrow \left(\mathbf{z}_{i1}^*,  y_{i1}^*\right)$, with $\left(\mathbf{z}_{i1}^*,  y_{i1}^*\right)$  has the largest Euclidean norm in $\mathbf{Z}_i$
\STATE Calculate $l\left(\mathbf{z}|\mathbf{Z}_{i1}^*\right)$ by \eqref{supp::algnew1}, for all $\mathbf{z} \in \mathbf{Z}_i/\mathbf{Z}_{i1}^*$
\FOR{$j=1$ to $m-1$}
\STATE $\mathbf{z}_{i(j+1)}^* \leftarrow \operatorname{arg}\underset{\mathbf{z} \in \mathbf{Z}_i/\mathbf{Z}_{ij}^*}{\min}\ l\left(\mathbf{z}|\mathbf{Z}_{ij}^*\right)$
\STATE $\left\{\mathbf{Z}_{i(j+1)}^*, \mathbf{Y}_{i(j+1)}^*\right\} \leftarrow \left\{\mathbf{Z}_{ij}^*, \mathbf{Y}_{ij}^*\right\} \bigcup \left\{\left(\mathbf{z}_{i(j+1)}^*, y_{i(j+1)}^*\right)\right\}$
\IF{$C_i \geqslant m^2$}
\STATE  Let $\kappa_{j} = C_i/j$
\ELSE
\STATE Let $\kappa_{j} = C_i / j^{r-1}$, where $r = \log(C_i) / \log(m)$
\ENDIF
\STATE $l\left(\mathbf{z}|\mathbf{Z}_{i(j+1)}^*\right) \leftarrow l\left(\mathbf{z}|\mathbf{Z}_{ij}^*\right) + l\left(\mathbf{z} \mid \mathbf{z}_{i(j+1)}^* \right)$,
for all $\mathbf{z} \in \left\{\mathbf{z}\right.$:  $\kappa_{j}$ points in  $\mathbf{Z}_i/\mathbf{Z}_{i(j+1)}^*$ with $\kappa_{j}$ smallest components in $ l\left(\mathbf{z}|\mathbf{Z}_{ij}^*\right) \}$
\ENDFOR

Let $\left\{\mathbf{Z}_{i}^*, \mathbf{Y}_{i}^*\right\} = \left\{\mathbf{Z}_{im}^*, \mathbf{Y}_{im}^*\right\}$
\end{algorithmic}
\end{algorithm}

\begin{remark}
The parameter $\kappa_{j}$ is set to keep $\kappa_{j}$ points in  $\mathbf{Z}_i/\mathbf{Z}_{i(j+1)}^*$ with $\kappa_{j}$ smallest components in $l\left(\mathbf{z}|\mathbf{Z}_{i(j)}^*\right)$. If $C_i \gg m$, say $C_i \geqslant m^2$, then we can eliminate a large number of unneeded candidate points, and set $\kappa_{j} = C_i/j$ to be much smaller than $C_i$; otherwise, we only eliminate a small portion and set $\kappa_{j} = C_i / j^{r-1}$ to be close to $n$, and $r = \log(C_i) / \log(m)$.
\end{remark}

\section{Estimating $\sigma_A^2$ and $\sigma_E^2$}
 \label{appD}
We use the method of moments introduced in \cite{gao2017efficient}.
Suppose that $n_{i}$ observations are sampled in the $i$th group. To estimate $\sigma_A^2$ and $\sigma_E^2$, the following U-statistics are used:
$$U_a=\frac{1}{2}\sum_{i,j,j'}\frac{1}{n_{i}}(\eta^*_{ij}-\eta^*_{ij'})^2=\frac{1}{2}\sum_{i,j,j'}\frac{1}{n_{i}}(e^*_{ij}-e^*_{ij'})^2,$$
$$U_e=\frac{1}{2}\sum_{i,j,i',j'}(\eta^*_{ij}-\eta^*_{i'j'})^2=\frac{1}{2}\sum_{i,j,i',j'}(a^*_i+e^*_{ij}-a^*_{i'}-e^*_{i'j'})^2,$$
where $\eta^*_{ij}=y^*_{ij}-\mathbf{x}_{ij}^{*T}\boldsymbol{\beta}=a^*_i+e^*_{ij}$, and $\sum_{i}$ denote $\sum_{i=1}^{R}$, $\sum_{j}$ denote $\sum_{j=1}^{n_i}$, so do $\sum_{i'}$ and $\sum_{j'}$.
Let $n = \sum_{i}n_{i}$. We have
\begin{align}
\mathrm{E}(U_a)&=\frac{1}{2}\sum_{i,j,j'}\frac{1}{n_{i}}[2\sigma_E^2(1-\mathbf{1}_{\{j=j'\}})] = \sigma_E^2(n-R),\notag\\
\mathrm{E}(U_e)&=\frac{1}{2}\sum_{i,j,i',j'}[2\sigma_A^2(1-\mathbf{1}_{\{i=i'\}})+2\sigma_E^2(1-\mathbf{1}_{\{i=i'\}}\mathbf{1}_{\{j=j'\}})]\notag\\
&=\sigma_A^2(n^2-\sum_{i}n_{i}^2)+\sigma_E^2(n^2-n),\notag
\end{align}
where $\mathbf{1}_{\{\cdot\}}$ is the indicator function.

Then, under conditions that the data size of pilot experiment $n \rightarrow \infty$, we have
\begin{align}
\mathrm{E}\begin{pmatrix}
U_a\\
U_e
\end{pmatrix}&=\begin{pmatrix}
0 &n-R \\
n^2-\sum_in_{i}^2 & n^2-n
\end{pmatrix}\begin{pmatrix}
\sigma_A^2\\
\sigma_E^2
\end{pmatrix}\notag\\
&=\begin{pmatrix}
n & 0\\
0 & n^2
\end{pmatrix}\begin{pmatrix}
0 &1-\frac{R}{n} \\
1-\frac{\sum_in_{i}^2}{n^2} & 1-\frac{1}{n}
\end{pmatrix}\begin{pmatrix}
\sigma_A^2\\
\sigma_E^2
\end{pmatrix}\notag\\
&=\begin{pmatrix}
n & 0\\
0 & n^2
\end{pmatrix}\begin{pmatrix}
0 &1 \\
1-\frac{\sum_in_{i}^2}{n^2} & 1
\end{pmatrix}(1+o(1))\begin{pmatrix}
\sigma_A^2\\
\sigma_E^2
\end{pmatrix}.\notag
\end{align}

Thus, the method of moments estimators of $\sigma_A^2$ and $\sigma_E^2$ can be expressed as
\begin{align}
  \hat{\sigma}_A^2=\frac{1}{n^2-\sum_in_{i}^2}\left(U_e - nU_a\right),\ \hat{\sigma}_E^2=\frac{1}{n}U_a.\label{supp::3}
\end{align}

In practice, we first use the selected subdata to obtain the ordinary least squares (OLS) estimators $\hat{\boldsymbol{\beta}}_{OLS}^*$, and then obtain $\hat{\sigma}_A^2$ and $\hat{\sigma}_E^2$ from (\ref{supp::3}), with $\hat{\eta}^*_{ij}=y^*_{ij}-\mathbf{x}_{ij}^{*T}\hat{\boldsymbol{\beta}}_{OLS}^*$. 

\end{appendix}
\clearpage
\bibliographystyle{apalike}
\bibliography{mybibfile}

\end{document}